\documentclass[aps,prl,twocolumn,superscriptaddress,groupedaddress]{revtex4}  
\usepackage{graphicx}  
\usepackage{dcolumn}   
\usepackage{bm}        
\usepackage{amssymb}   
\usepackage{lingmacros}
\usepackage{tree-dvips}
\usepackage{mathtools}
\usepackage[utf8]{inputenc}
\usepackage{graphicx}
\usepackage[rightcaption]{sidecap}
\usepackage{float}
\usepackage{wrapfig}
\usepackage{amsmath}
\usepackage{parskip}
\usepackage{pdfpages}
\usepackage[symbol]{footmisc}

\usepackage{color}

\begin{document}
	\setlength{\parskip}{0pt}
	\setlength{\parindent}{1em}

	%
	\title{Enhancing nonlinear damping by parametric-direct internal resonance}
	
	
	\author{Ata Keşkekler\footnote[1]{Corresponding authors:\\Ata Keşkekler $<$a.keskekler-1@tudelft.nl$>$\\ Farbod Alijani $<$f.alijani@tudelft.nl$>$}}
\affiliation{Department of Precision and Microsystems Engineering, TU Delft, The Netherlands}
	
	\author{Oriel Shoshani}
	\affiliation{Department of Mechanical Engineering, Ben-Gurion University of Negev, Israel}
	
		\author{Martin Lee}
	\affiliation{Kavli Institute of Nanoscience, TU Delft, The Netherlands}
	
	\author{Herre S. J. van der Zant}
	\affiliation{Kavli Institute of Nanoscience, TU Delft, The Netherlands}

	\author{Peter G. Steeneken}
	\affiliation{Department of Precision and Microsystems Engineering, TU Delft, The Netherlands}
	\affiliation{Kavli Institute of Nanoscience, TU Delft, The Netherlands}

	\author{Farbod Alijani$^\ast$}
	\affiliation{Department of Precision and Microsystems Engineering, TU Delft, The Netherlands}
	
	\date{\today}

	\begin{abstract}
Mechanical sources of nonlinear damping play a central role in modern physics, from solid-state physics 
to thermodynamics. The microscopic theory of mechanical dissipation [ M. I . Dykman, M. A. Krivoglaz, Physica Status Solidi (b) 68, 111 (1975)] suggests that nonlinear damping of a resonant mode can be strongly enhanced when it is coupled to a vibration mode that is close to twice its resonance frequency. To date, no experimental evidence of this enhancement has been realized. In this letter, we experimentally show that nanoresonators driven into parametric-direct internal resonance provide supporting evidence for the microscopic theory of nonlinear dissipation. By regulating the drive level, we tune the parametric resonance of a graphene nanodrum over a range of 40-70 MHz to reach successive two-to-one internal resonances, leading to a nearly two-fold increase of the nonlinear damping. Our study opens up an exciting route towards utilizing modal interactions and parametric resonance to realize resonators with engineered nonlinear dissipation over wide frequency range.

	\end{abstract}
	
	\maketitle
	In nature, from macro to nano scale, dynamical systems evolve towards thermal equilibrium while exchanging energy with their surroundings. Dissipative mechanisms that mediate this equilibration, convert energy from the dynamical system of interest to heat in an environmental bath. This process can be extremely intricate, nonlinear, and in most cases hidden behind the veil of linear viscous damping, which is merely an approximation valid for small amplitude oscillations.
	
	In the last decade, nonlinear dissipation has attracted much attention in the study of mechanical systems with applications that span nanomechanics \cite{bachtoldfirst}, materials science \cite{Amabili2019}, biomechanics \cite{amabili2020nonlinear}, thermodynamics \cite{midtvedt2014fermi}, and quantum information \cite{leghtas2015confining}. It has been shown that the nonlinear dissipation process in these wide range of applications follows the empirical force model $F_d=-\tau_{\rm nl1} x^2 \dot x$ where $\tau_{\rm nl1}$ is the nonlinear damping coefficient, $x$ is the displacement and $\dot x$ velocity. To date, the physical mechanism from which this empirical damping force originates has remained ambiguous, with a diverse range of phenomena being held responsible including viscoelasticity \cite{viscoZaitsev}, phonon-phonon interactions \cite{croyPhonon}, Akheizer relaxation \cite{atalaya2016nonlinear}, and mode coupling \cite{guttinger2017energy}. The fact that nonlinear damping can stem from multiple origins simultaneously, makes isolating one route from the others a daunting task, especially since the nonlinear damping coefficient $ \tau_{\rm nl1}$ is perceived to be a fixed parameter that unlike stiffness \cite{song2012stamp,sajadi2017experimental,lee2019sealing}, quality factor \cite{miller2018effective}, and nonlinear stiffness \cite{weber2014coupling,samanta2018tuning,yang2020persistent}, cannot be tuned easily.
	
	Amongst the different mechanisms that affect nonlinear damping, intermodal coupling is particularly interesting, as it can be enhanced near internal resonance (IR), a special condition at which the ratio of the resonance frequencies of the coupled modes is a rational number \cite{nayfnonosc}. This phenomenon has frequently been observed in nano/micro-mechanical resonators \cite{herreclamped,antonio2012frequency,eichler2012strong,chen2017direct,shoshani2017anomalous,czaplewski2018bifurcation,czaplewski2019bifurcation,yang2019spatial,houri2019multimode,van2010amplitude}. At internal resonance, modes can interact strongly even if their nonlinear coupling is relatively weak. Interestingly, internal resonance is closely related to the effective stiffness of resonance modes, and can therefore be manipulated by careful engineering of the geometry of mechanical systems, their spring hardening nonlinearity \cite{herredirectpar}, and electrostatic spring softening \cite{van2010amplitude}. Internal resonance also finds its route in the microscopic theory of dissipation proposed back in 1975, where it was hypothesized to lead to a significantly shorter relaxation time if there exists a resonance mode in the vicinity of twice the resonance frequency of the driven mode in the density of states \cite{dykman1975spectral}.
	
	
	In this letter, we demonstrate that nonlinear damping of graphene nanodrums can be strongly enhanced by parametric-direct internal resonance, providing supporting evidence for the microscopic theory of nonlinear dissipation \cite{dykman1975spectral,atalaya2016nonlinear}. To achieve this, we bring the fundamental mode of the nanodrum into parametric resonance at twice its resonance frequency, allowing it to be tuned over a wide frequency range from 40-70 MHz. 
	We extract the nonlinear damping as a function of the parametric drive level, and observe that it increases as much as 80 \% when the frequency shift of the parametric resonance brings it into internal resonance with a higher mode. By comparing the characteristic dependence of the nonlinear damping coefficient on parametric drive to a theoretical model, we confirm that internal resonance can be held accountable for the significant increase in nonlinear damping.

\begin{figure}
		\centering
		\includegraphics[width=1\linewidth]{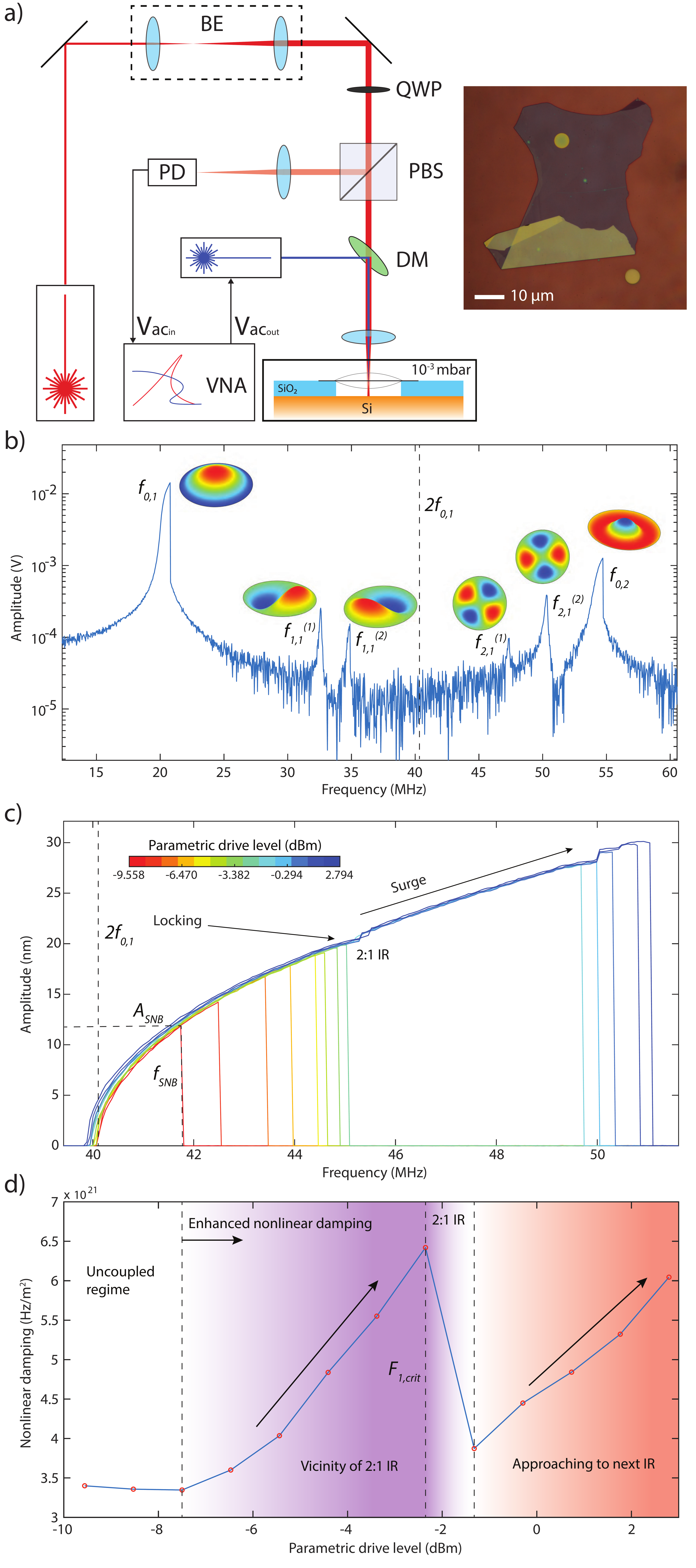}
		\caption{Nonlinear dynamic response of a graphene nanodrum near 2:1 internal resonance. (a) Fabry–Pérot interferometry  with thermomechanical actuation and microscope image of the graphene. Experiments are performed in vacuum at $10^{-3}$ mbar. In the figure; BE: Beam expander, QWP: Quarter wave plate, PBS: Polarized beam splitter, PD: Photodiode, DM: Dichroic mirror, VNA: Vector network analyzer. (b) Direct frequency response curve of the device, showing multiple resonances (Drive level = -12.6 dBm). The mode shapes are simulated by Comsol. 
			(c) Parametric resonance curves, driven at twice the detection frequency. 
			(d) Variation of the nonlinear damping $\tau_{\rm nl1}$ as a function of drive $F_1$.}
		\label{fig:firstbiggraphv2}
	\end{figure}

	Experiments are performed on a 10 nm thick multilayer graphene nanodrum with a diameter of 5 $\mu$m, that is transferred over a cavity etched in a layer of Si$\rm O_2$ with a depth of 285 nm. We use a power modulated blue laser ($\lambda= $ 488 nm) to thermomechanically actuate the nanodrum. We then read-out the motion using a red laser ($\lambda= $ 633 nm) whose reflected intensity is modulated by the motion of the nanodrum in a  Fabry–P\'erot etalon formed by the graphene and the Si back mirror (Fig. \ref{fig:firstbiggraphv2}a). The reflected red laser intensity from the center of the drum is detected using a photodiode, whose response is read by the same Vector Network Analyzer (VNA) that modulates the blue laser. The measured VNA signal is then converted to displacement in nanometers using a nonlinear optical calibration method \cite{robincalib} (see Supplemental Material \cite{SM} I).
	
	By sweeping the drive frequency we obtain the frequency response of the nanodrum in which multiple directly driven resonance modes can be identified (Fig. \ref{fig:firstbiggraphv2}b). We find the fundamental axisymmetric mode of vibration at $f_{\rm 0,1}$=20.1 MHz and several other modes, of which the two modes, at $f^{(1)}_{\rm 2,1}$=47.4 MHz and $f^{(2)}_{\rm 2,1}$=50.0 MHz, are of particular interest. This is because, to study the effect of internal resonance on nonlinear damping, we aim to achieve a two-to one (2:1) internal resonance by parametrically driving the fundamental mode, such that it coincides with one of the higher frequency modes. The frequency ratios $f^{(1)}_{\rm 2,1}/f_{\rm 0,1}\approx 2.3$ and $f^{(2)}_{\rm 2,1}/f_{\rm 0,1}\approx 2.4$ are close to the factor 2, however additional frequency tuning is needed to reach the 2:1 internal resonance condition.
	
	The parametric resonance can be clearly observed by modulating the tension of the nanodrum at frequency $\omega_F$ with the blue laser while using a frequency converter in the VNA to measure the amplitude at $\omega_F/2$ as shown in Fig. \ref{fig:firstbiggraphv2}c. By increasing the parametric drive, we observe a Duffing-type geometric nonlinearity over a large frequency range, such that the parametrically driven fundamental resonance can be tuned across successive 2:1 internal resonance conditions with modes $f^{(1)}_{\rm 2,1}$ and $f^{(2)}_{\rm 2,1}$, respectively.
	
	In Fig. \ref{fig:firstbiggraphv2}c we observe that the parametric resonance curves follow a common response until they reach the saddle-node bifurcation frequency $f_{\rm SNB}$ above which the parametric resonance curve reaches its peak amplitude $A_{\rm SNB}$ and drops down to low 
	amplitude. We note that the value of $A_{\rm SNB}$ can be used to determine the degree of nonlinear damping \cite{lifshitz2008nonlinear}. Therefore, to extract the nonlinear damping coefficient $\tau_{\rm nl1}$ of mode $f_{\rm 0,1}$ from the curves in Fig. \ref{fig:firstbiggraphv2}c, we use the following single degree-of-freedom (DoF) model to describe the system dynamics:
	\begin{equation}
	\ddot x_1+\omega_1^2x_1+\gamma x^3=F_{1}x_1\cos(\omega_{F}t)-2\tau_{1}\dot x_1-2\tau_{\rm nl1}x_{1}^2\dot x_1,
	\label{eq:single_mode}
	\end{equation}
	in which $\omega_1=2 \pi f_{\rm 0,1}$ is the eigenfrequency of the axisymmetric mode of the nanodrum, $\gamma$ is its Duffing constant and $F_1$ and $\omega_F$ are the parametric drive amplitude and frequency, respectively. Moreover, $2\tau_1=\omega_1/Q$ is the linear damping coefficient, with $Q$ being the quality factor, and  $\tau_{\rm nl1}$ is the nonlinear damping term of van der Pol type that prevents the parametric resonance amplitude $A_{\rm SNB}$ from increasing to infinity \cite{lifshitz2008nonlinear, robinparam} at higher driving frequencies since $|A_{\rm SNB}|^2\propto (2 F_1 Q-4)/\tau_{\rm nl1}$.
	
	To identify the parameters governing the device dynamics from the measurements in Fig. \ref{fig:firstbiggraphv2}c, we use Eq. (\ref{eq:single_mode}) and obtain good fits of the parametric resonance curves using $\tau_{\rm nl1}$ and $\gamma$ as fit parameters (see Supplemental Material \cite{SM} IV).
	
	As we gradually increase the drive level, $f_{\rm SNB}$ increases until it reaches the vicinity of the internal resonance, where we observe an increase in $\tau_{\rm nl1}$ (Fig. \ref{fig:firstbiggraphv2}d).  Whereas $f_{\rm SNB}$ increases with parametric drive $F_{1}$, Fig. \ref{fig:firstbiggraphv2}c shows that its rate of increase $\frac{{\rm d}f_{\rm SNB}}{{\rm d}F_1}$ slows down close to $f^{(1)}_{\rm 2,1}$, locking the saddle-node-bifurcation frequency when $f_{\rm SNB}\approx$ 45 MHz. At the same time, $\tau_{\rm nl1}$ increases significantly at the associated parametric drive levels, providing the possibility to tune nonlinear damping up to two-folds by controlling $F_{1}$, as seen in Fig. \ref{fig:firstbiggraphv2}d. We note that a similar enhancement of damping was also observed in \cite{guttinger2017energy} for a graphene nanodrum undergoing 3:1 internal resonance. However, in that case,  the mechanism is more intricate and leads to higher (quintic) nonlinear damping that comes into play at relatively large amplitude oscillations \cite{shoshani2017anomalous}.
	
	Fig. \ref{fig:firstbiggraphv2}c also shows that above a certain critical parametric drive level $F_{\rm 1,crit}$, the frequency locking barrier at $f_{\rm SNB}\approx$ 45 MHz is broken and $f_{\rm SNB}$ suddenly jumps to a higher frequency ($\approx$ 5 MHz higher), and a corresponding larger $A_{\rm SNB}$. We label this increase in the rate $\frac{{\rm d}f_{\rm SNB}}{{\rm d}F_1}$ by ``surge" in Fig. \ref{fig:firstbiggraphv2}c, where an abrupt increase in the amplitude-frequency response is observed to occur above a critical drive level $F_{\rm 1,crit}$. Interestingly, even above $F_{\rm 1,crit}$ a further increase in $\tau_{\rm nl1}$ is observed  with increasing drive amplitude, indicating that a similar frequency-locking  occurs when the parametric resonance peak reaches the second internal resonance at $f_{\rm SNB}\approx f^{(2)}_{\rm 2,1}$.
	
	Although the 1 DoF model in Eq. (\ref{eq:single_mode}) can capture the response of the parametric resonance, it can only do so by introducing a non-physical drive level dependent nonlinear damping coefficient $\tau_{\rm nl1}(F_1)$ (Fig. \ref{fig:firstbiggraphv2}d). Therefore, to study the physical origin of our observation, we extend the model by introducing a second mode whose motion is described by generalized coordinate $x_2$. Moreover, to describe the coupling between the interacting modes at the 2:1 internal resonance, we use the single term coupling potential $U_{\rm cp}=\alpha x_1^2 x_2$ (see Supplemental Material \cite{SM} II). The coupled equations of motion in the presence of this potential become:
	
	\begin{small}
		\begin{gather}
		\ddot x_1+\omega_1^2 x_1+\gamma x_1^3+\frac{\partial U_{\rm cp}}{\partial x_1}=F_{1}x_{1}\cos(\omega_F t)-2\tau_{1} \dot
		x_1-2\tau_{nl1}x_1^2 \dot x_1,\nonumber\\
		\ddot x_2+\omega_2^2 x_2+\frac{\partial U_{\rm cp}}{\partial x_2}=F_{2}\cos(\omega_F t)-2\tau_2 \dot x_2.
		\label{eq:gov_eqs}
		\end{gather}
	\end{small}
	
	\noindent
	The 2 mode model describes a parametrically driven mode with generalized coordinate $x_1$ coupled to $x_2$ that has  eigenfrequency $\omega_2=2 \pi f^{(1)}_{\rm 2,1}$, damping ratio $\tau_2$, and is directly driven by a harmonic force with magnitude $F_2$. 
	
	To understand the dynamics of the system observed experimentally and described by the model in Eq. (\ref{eq:gov_eqs}), it is convenient to switch to the rotating frame of reference by transforming $x_1$ and $x_2$ to complex amplitude form (see Supplemental Material \cite{SM} III). This transformation reveals a system of equations that predicts the response of the resonator as the drive parameters ($F_1,~F_2$, and $\omega_F$) are varied. Solving the coupled system at steady-state yields the following algebraic equation for the amplitude $a_1$ of the first mode:
	\begin{align}
	&\left[\tau_{1}+(\tau_{\rm nl1}+\tilde\alpha^2\tau_2)\frac{a_1^2}{4}\right]^2+\left[\Delta\omega_{1}-\left(\frac{3\gamma}{\omega_F}+\tilde\alpha^2\Delta\omega_{2}\right)\frac{a_1^2}{4}\right]^2\nonumber\\
	&=\frac{1}{4\omega_F^2}\Big[F_1^2+\tilde\alpha^2(F_2^2+2\omega_F\Delta\omega_{2}F_1 F_2/\alpha)\Big],
	\label{eq:res_curve}
	\end{align}
	where $\Delta\omega_1=\omega_F/2-\omega_1$ and $\Delta\omega_2=\omega_F-\omega_2$ are the frequency detuning from the primary and the secondary eigenfrequencies, and $\tilde\alpha^2=\alpha^2/[\omega_F^2(\tau_2^2+\Delta\omega_2^2)]$ is the rescaled coupling strength. Essentially, the first squared term in (\ref{eq:res_curve}) captures the effect of damping on the parametric resonance amplitude $a_1$, the second term captures the effect of nonlinear coupling on the stiffness and driving frequency, and the term on the right side is the effective parametric drive. From the rescaled coupling strength $\tilde\alpha$ and Eq.  (\ref{eq:res_curve}) it can be seen that the coupling $\tilde\alpha^2$ shows a large peak close to the 2:1 internal resonance where $|\Delta\omega_2|\approx 0$. 
	Interestingly, this shows that the 2 mode model can account for an increase in the effective nonlinear damping parameter $\tau_{nl\rm eff}=\tau_{\rm nl1}+\tilde\alpha^2\tau_2$ near internal resonance, in accordance with the observed peak in $\tau_{\rm nl1}$ obtained from the experimental fits with the 1 DoF model in Fig. \ref{fig:firstbiggraphv2}d.  
	
	
	The 2 mode model of Eq. (\ref{eq:res_curve}) allows us to obtain good fits of the parametric resonance curves in Fig. \ref{fig:firstbiggraphv2}b, with a constant $\tau_{\rm nl1}\approx 3.4\times10^{21}$ (Hz/m$^2)$  determined far from internal resonance and a single coupling strength $\alpha= 2.2\times10^{22}$ (Hz$^2$/m) which intrinsically accounts for the variation of $\tau_{\rm nleff}$ near internal resonance. These fits can be found in Supplemental Material \cite{SM} V, and demonstrate that the 2 mode model is in agreement with the experiments for constant parameter values, without requiring drive level dependent fit parameters.
	\begin{figure}
		\centering
		\includegraphics[width=1\linewidth]{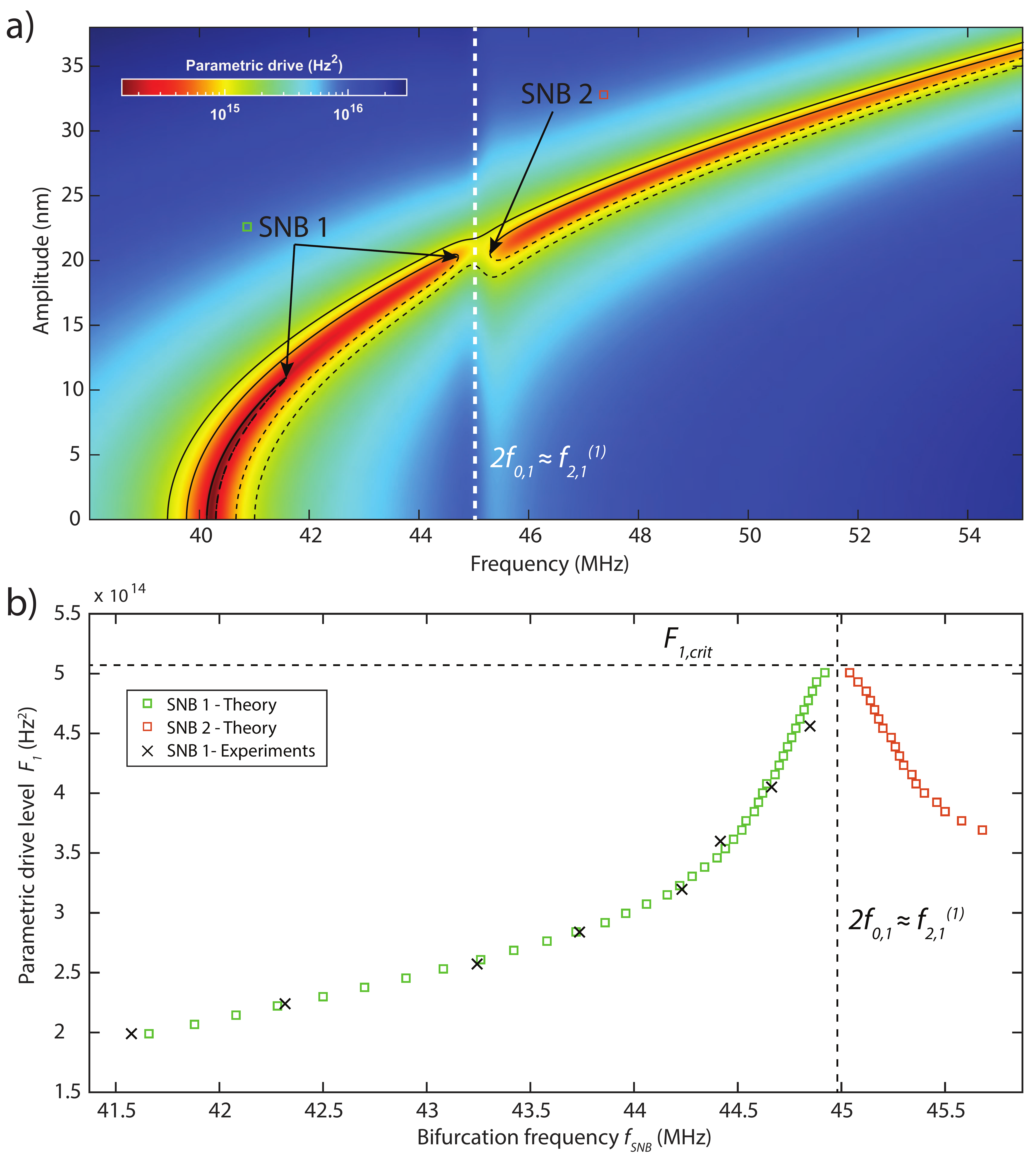}
		\caption{(a) Color map of the analytical model response curves obtained by using the fitted parameters from experiments. (b) The underlying route of the amplitude-frequency surge is revealed by tracing the evolution of saddle-node bifurcations (SNB1 and SNB2) of the parametric resonance curves.}
		\label{fig:secondbig}
	\end{figure}
	
	To understand the physics associated with the frequency-locking and amplitude-frequency surge, we use the experimentally extracted fit parameters from the 2 mode model and numerically generate parametric resonance curves using Eq. (\ref{eq:res_curve}) for a large range of drive amplitudes (see Fig. \ref{fig:secondbig}a).
	We see that for small drive levels, an upward frequency sweep will follow the parametric resonance curve and then will lock and jump-down at the first saddle-node bifurcation (SNB1) frequency, that lies close to $f_{\rm SNB} \approx f^{(1)}_{\rm 2,1}$. At higher parametric drive levels, the parametric resonance has a stable path to traverse the internal resonance towards a group of stable states at higher frequencies.
	
	A more extensive investigation of this phenomenon can be carried out by performing bifurcation analysis of the steady-state solutions (see Supplementary Material \cite{SM} III). 
	\begin{figure}
		\centering
		\includegraphics[width=1\linewidth]{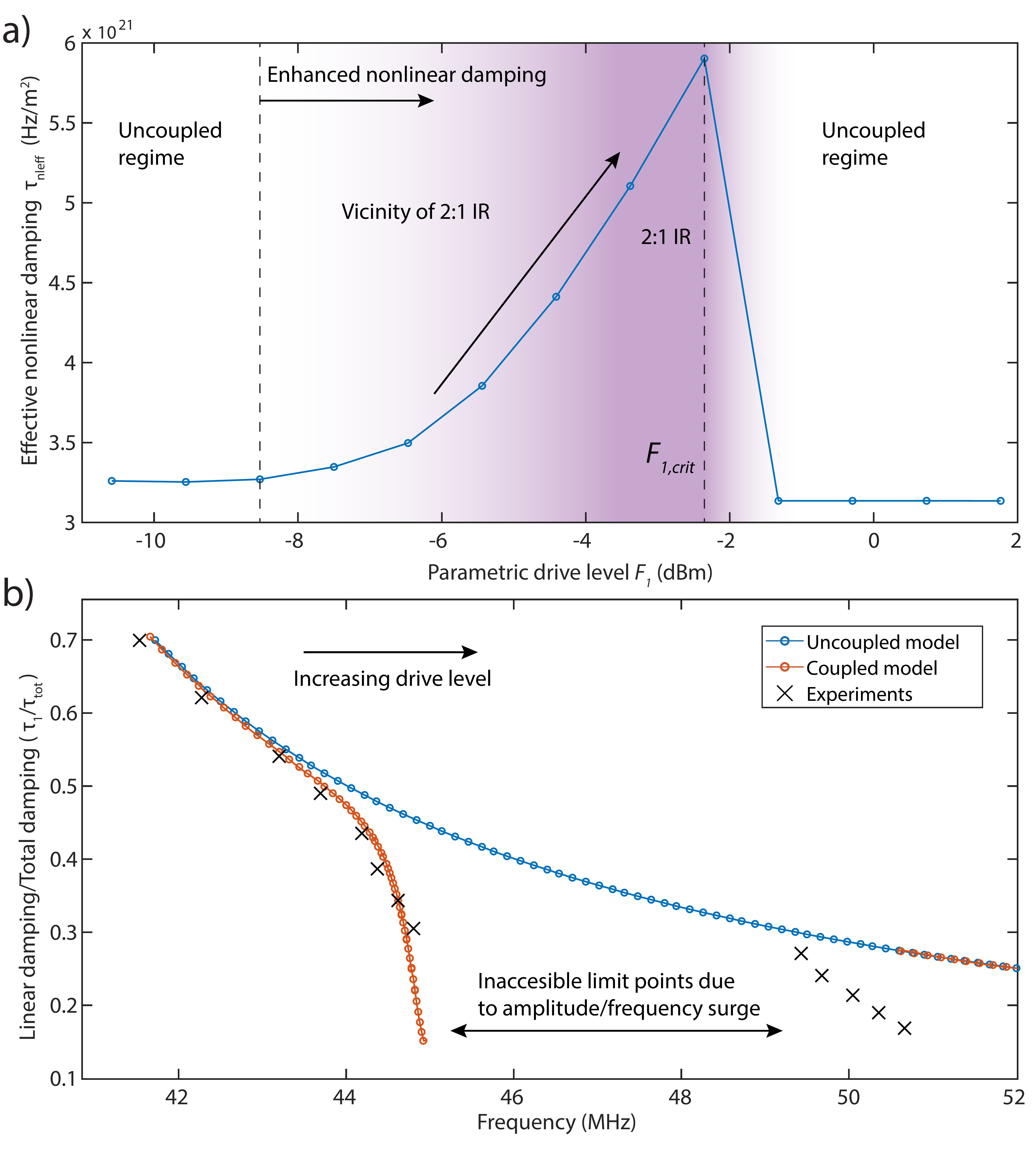}
		\caption{(a) Variation of the effective nonlinear damping parameter ($\tau_{\rm nl\rm eff}$) with respect to parametric drive. The $\tau_{\rm nl\rm eff}$ is obtained by fitting the numerically generated curves of Fig.\ref{fig:secondbig}a as the fit parameter. (b) Comparison of the ratio between linear damping and total damping from experiments and theoretical calculations.} 
		\label{fig:bifs}
	\end{figure}
	The bifurcation analysis reveals two saddle-node bifurcations near the singular region of the internal resonance, one at the end of the first path (SNB1) and another at the beginning of the second path (SNB2) (Fig. \ref{fig:secondbig}b). As the drive amplitude increases, the bifurcation pair starts to move towards each other until they annihilate one another to form a stable solution at the connecting point, which we labeled as "surge". It is also possible to observe that the rate at which saddle-node pairs approach each other dramatically drops near the internal resonance condition, demonstrating the "locking" which we also observed in the experiments.
	
	To check how closely the 2 mode model captures the variation of $\tau_{\rm nl1}$ close to the internal resonance condition, we follow a reverse path, and fit the numerically generated resonance curves of Fig. \ref{fig:secondbig}a using the 1 DoF model of Eq. (\ref{eq:res_curve}) with $\tau_{\rm nl1}$ as the fit parameter. In this way, we track the variation of $\tau_{\rm nl1}$ in the 1 DoF model with the parametric drive $F_1$, similar to what we observed experimentally and reported in Fig. \ref{fig:firstbiggraphv2}c. The result of this fit is shown in  Fig. \ref{fig:bifs}a, where a similar anomalous change of nonlinear damping is obtained for the 2 mode model.
	
	The variation of nonlinear damping affects the total damping (sum of linear and nonlinear dissipation) of the resonator too. It is of interest to study how large this effect is. In Fig. \ref{fig:bifs}b we report the variation in the ratio of the linear damping $\tau_1$ and the amplitude-dependent total damping $\tau_{\rm tot}=(\omega_1/Q + 0.25 \tau_{nl\rm eff} |x_{1}|^2)$ \cite{lifshitz2008nonlinear} in the spectral neighborhood of $f^{(1)}_{\rm 2,1}$, and observe a sudden decrease in the vicinity of internal resonance. This abrupt change in the total damping is well captured by the 2 mode model. With the increase in the drive amplitude, $\tau_1/\tau_{\rm tot}$ values of this model though, deviate from those of the experiments due to a subsequent internal resonance at $f^{(2)}_{\rm 2,1}/f_{\rm 0,1} \approx 2.4$ that is not included in our theoretical analysis. The dependence of $\tau_1/\tau_{\rm tot}$ on frequency shows that near internal resonance the total damping of the fundamental mode increases nearly by 80\%.

\begin{figure}
		\centering
		\includegraphics[width=1\linewidth]{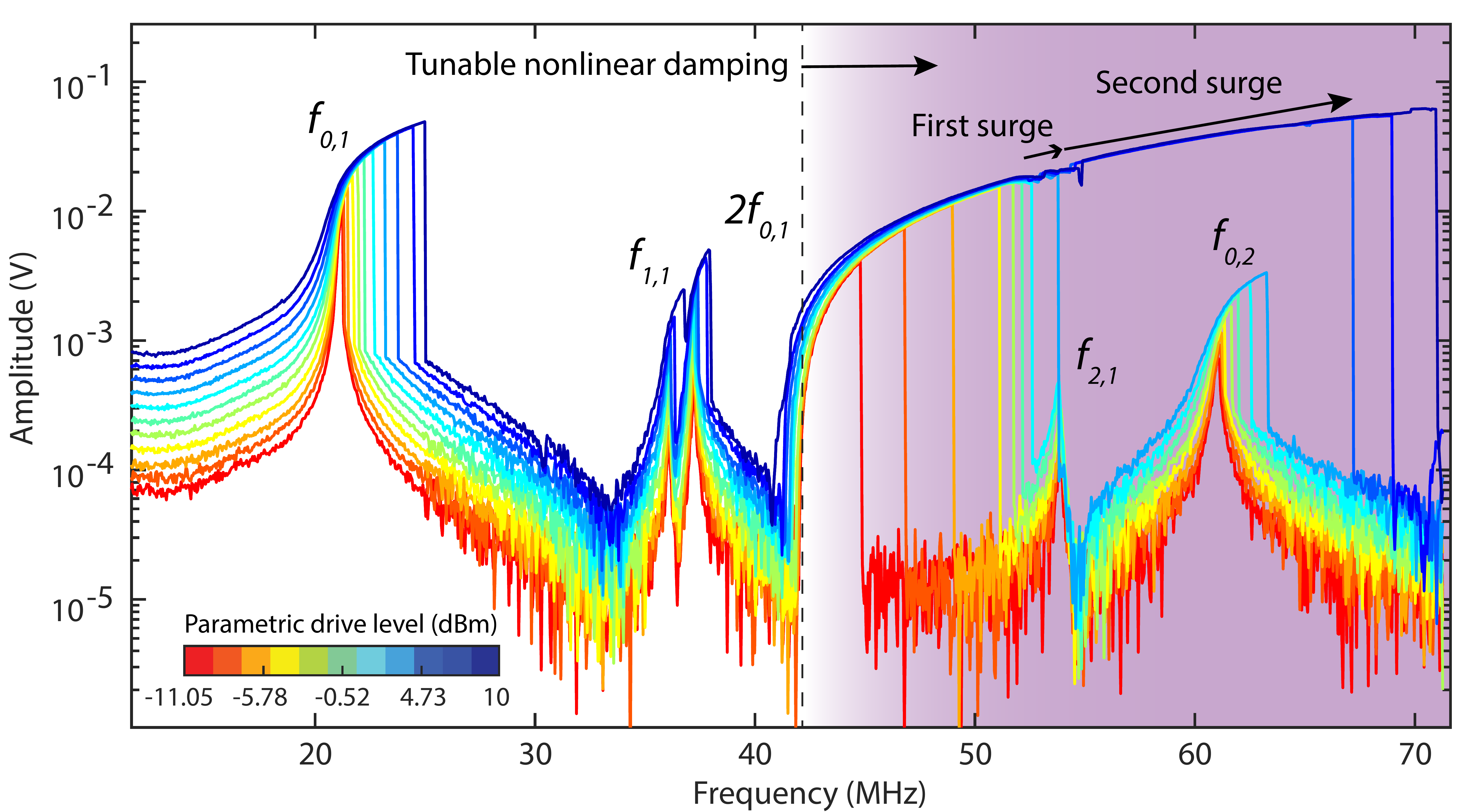}
		\caption{Frequency response measurements at high drive powers, pushing the  parametric resonance response into successive internal resonances. The arrows in the figure show successive amplitude-frequency surges.}
		\label{fig:m1e1measurement}
	\end{figure}

	When increasing the blue laser power and modulation, the parametrically actuated signal is also observed in the direct detection mode (like in Fig. 1b) due to optical readout nonlinearities \cite{robincalib}. As a result a superposition of Fig. 1b and 1c is obtained, as shown in Fig. \ref{fig:m1e1measurement}. 
	In this configuration we achieve a frequency shift in $f_{\rm SNB}$ from 40-70 MHz, corresponding to as much as 75 \% tuning of the mechanical motion frequency. This large tuning can increase the number of successive internal resonances that can be reached even further, to reach modal interactions between the parametric mode $f_{\rm 0,1}$ and direct modes $f^{(2)}_{\rm 2,1}$ and $f_{\rm 0,2}$ (see Fig. \ref{fig:m1e1measurement}). As a result, multiple amplitude-frequency surges can be detected in the large frequency range of 30 MHz over which nonlinear damping coefficient can be tuned.
	
	In summary, we study the enhancement of nonlinear damping in a graphene nanomechanical resonator, where the fundamental mode is parametrically driven to interact with a higher mode. When the system is brought near a 2:1 internal resonance, a significant increase in nonlinear damping is observed. In addition, the rate of increase of the parametric resonance frequency reduces in a certain locking regime,  stabilizing the values of $f_{\rm SNB}$ and $A_{\rm SNB}$, which could potentially aid frequency noise reduction \cite{antonio2012frequency}. 
	Interestingly, as the drive level is further increased beyond the critical level $F_{\rm 1,crit}$, this locking barrier is broken, resulting in a surge in $f_{\rm SNB}$ and amplitude of the resonator. These phenomena were studied experimentally, and could be accounted for using a 2 mode theoretical model. The described mechanism can isolate and differentiate mode coupling induced nonlinear damping from other dissipation sources,  and sheds light on the origins of nonlinear dissipation in nanomechanical resonators. It also provides a way to controllably tune nonlinear damping which complements existing methods for tuning linear damping \cite{miller2018effective}, linear stiffness \cite{song2012stamp,sajadi2017experimental,lee2019sealing} and nonlinear stiffness\cite{weber2014coupling,samanta2018tuning, yang2020persistent}, 
	extending our toolset to adapt and study the rich nonlinear dynamics of nanoresonators. 
	


	\begin{acknowledgments}
		We thank Prof. Marco Amabili for fruitful discussions about nonlinear damping. The research leading to these results received funding from European Union’s Horizon 2020 research and innovation program under Grant Agreement 802093 (ERC starting grant ENIGMA). O.S. acknowledges support for this work from the United States -- Israel Binational Science Foundation under Grant No. 2018041.  P.G.S. and H.S.J.v.d.Z. acknowledge funding from the European Union’s Horizon 2020 research and innovation program under grant agreement numbers 785219 and 881603 (Graphene Flagship).
	\end{acknowledgments}

\onecolumngrid
\clearpage


\section*{Supplementary material}

# Supplemental Material: Enhancing nonlinear damping by parametric-direct internal resonance

Ata Keşkekler,[1] Oriel Shoshani,[2] Martin Lee,[3] Herre. S. J. van der Zant,[3] Peter G. Steeneken,[1, 3] and Farbod Alijani[1]

[1]*Department of Precision and Microsystems Engineering, TU Delft, The Netherlands*
[2]*Department of Mechanical Engineering, Ben-Gurion University of Negev, Israel*
[3]*Kavli Institute of Nanoscience, TU Delft, The Netherlands*

## SUPPLEMENTAL MATERIAL I. AMPLITUDE CALIBRATION

We measure the response of the graphene membrane in Volts from a Fabry–Pérot interferometer and convert the readings to nanometers using the nonlinear optical transduction technique presented in [1]. We note that for thin membranes and for high back-mirror reflectively, the reflected intensity $I(t)$ of the optical read-out can be approximated by

$$I(t) = A + B \cos\left(4\pi \frac{g + \bar{x}(t)}{\lambda}\right), \tag{1}$$

where $A$ and $B$ are constants, $\bar{x}(t) = x\sin(\omega_F t)$ is the membrane displacement, $g$ is the cavity depth, and $\lambda$ is the wave-length of the light used for measurement. The reflected intensity $I(t)$ in Eq. (1) is a nonlinear function of the membrane displacement, and therefore, the read-out of a monochromatic signal will contain higher-order harmonics . The amplitude of these harmonics can be obtained from a Fourier expansion of the intensity $I(t) = \sum I_{n\omega_F} \sin(n\omega_F t)$, and harmonic balancing of Eq. (1). This will lead to the following relations for the first ($I_{1\omega_F}$) and the third ($I_{3\omega_F}$) harmonics of the motion [1]:

$$I_{1\omega_F} = -B\eta x \sin(\eta g) + \frac{1}{8}Bx^3\eta^3 \sin(\eta g), \tag{2}$$

$$I_{3\omega_F} = -\frac{1}{24}Bx^3\eta^3 \sin(\eta g), \tag{3}$$

in which $\eta = 4\pi/\lambda$. Taking the ratio of the two harmonics, one can find the motion amplitude as follows:

$$x = \frac{2\sqrt{6}}{\eta} \sqrt{\frac{I_{3\omega_F}/I_{1_F}}{1 + 3I_{3\omega_F}/I_{1\omega_F}}}. \tag{4}$$

We can then obtain the linear transduction coefficient $C_{conv} = x/I_{1\omega_F}$ by averaging multiple data points on the nonlinear resonance curves in Fig. 1, in order to calibrate the response where $I_{3\omega_F}$ is below the noise level. The value we obtain for our experiments is $C_{conv} = 7.92 \times 10^{-7}$ (m/V). We also note that, above certain amplitudes, saturation of the resonance frequency curves become apparent for which the nonlinear transduction error becomes significant. We correct for this using [1]

$$x_{nl} = \left(1 + \frac{1}{8}x^2\eta^2\right)x. \tag{5}$$

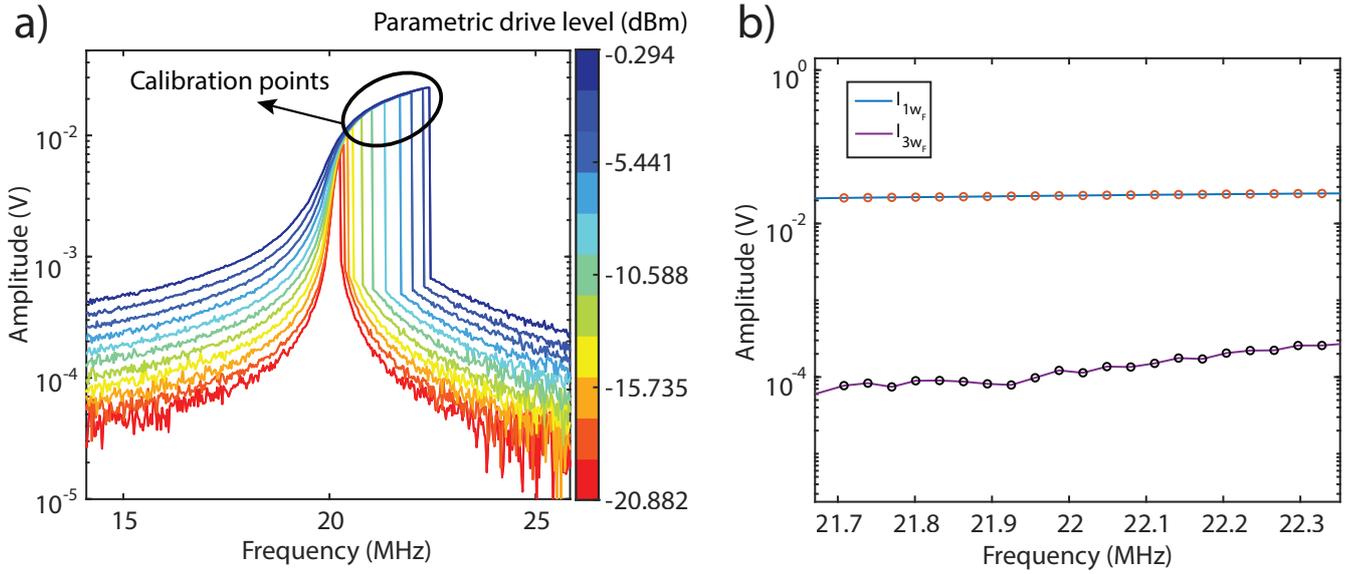

FIG. 1: (a) Nonlinear optical effect observed in the Duffing response has been used for amplitude calibration. (b) First and third harmonics arising from optical nonlinearity. The ratio of these harmonics are used to determine oscillation amplitude.

## SUPPLEMENTAL MATERIAL II. NORMAL FORM OF THE EQUATIONS OF MOTION

In the absence of damping and external forces, the geometrically nonlinear equations of motion for plates and membranes can be written in the following discretized form [2]

$$\ddot{x}_k + \omega_k^2 x_k + \sum_n \sum_p \alpha_{np}^{(k)} x_n x_p + \sum_n \sum_p \sum_l \gamma_{npl}^{(k)} x_n x_p x_l = 0, \quad \forall k \in \mathbb{N}, \tag{6}$$

where $\omega_k$ is the eigenfrequency of the $k^{th}$ mode $x_k$, and $\alpha_{np}^{(k)}$ and $\gamma_{npl}^{(k)}$ are quadratic and cubic nonlinear terms, respectively. For a two mode system $(k = 2)$, the nonlinear equations become

$$\ddot{x}_1 + \omega_1^2 x_1 + \alpha_{11}^{(1)} x_1^2 + \alpha_{12}^{(1)} x_1 x_2 + \alpha_{22}^{(1)} x_2^2 + \gamma_{111}^{(1)} x_1^3 + \gamma_{112}^{(1)} x_1^2 x_2 + \gamma_{122}^{(1)} x_1 x_2^2 + \gamma_{222}^{(1)} x_2^3 = 0, \tag{7}$$

$$\ddot{x}_2 + \omega_2^2 x_1 + \alpha_{11}^{(2)} x_1^2 + \alpha_{12}^{(2)} x_1 x_2 + \alpha_{22}^{(2)} x_2^2 + \gamma_{111}^{(2)} x_1^3 + \gamma_{112}^{(2)} x_1^2 x_2 + \gamma_{122}^{(2)} x_1 x_2^2 + \gamma_{222}^{(2)} x_2^3 = 0. \tag{8}$$

Note that many of the nonlinear terms in Eq. (7) and (8) are non-resonant, and hence, can be eliminated via a normal form transformation [3]. To recover the resonant terms (which cannot be eliminated from the normal form) in a 2:1 internal resonance condition ($\omega_2 \simeq 2\omega_1$), we assume harmonic motion of the form $x_1 \approx \cos(\omega_1 t)$ and $x_2 \approx \cos(2\omega_1 t)$ as a first approximation. Inserting these relations in Eq. (7) reveals that the terms $x_1^3 \approx \frac{3}{4}\cos(\omega_1 t) + \frac{1}{4}\cos(\omega_1 t)$ and $x_1 x_2 \approx \frac{1}{2}(\cos(\omega_1 t) + \cos(3\omega_1 t))$ in the first equation of motion are trivially resonant. The same holds for the term $x_1^2 \approx \approx \frac{1}{2}(1 + \cos(2\omega_1 t))$, which can be viewed as a resonant term for Eq. (8). Furthermore, in order to obtain the most simple model for the considered dynamical system, we neglect the contribution of the dispersive coupling terms $x_1 x_2^2$ in Eq. (7) and $x_2^2 x_1$ in Eq. (8) (which only shift the resonance frequency of each mode in amount that is proportional to the amplitude square of the other mode), and the Duffing nonlinearity of the second mode (which is assumed to operate below the Duffing threshold). Therefore, the governing equations of motion reduce to

$$\ddot{x}_1 + \omega_1^2 x_1 + \alpha_{12}^{(1)} x_1 x_2 + \gamma_{111}^{(1)} x_1^3 = 0, \tag{9}$$

$$\ddot{x}_2 + \omega_2^2 x_1 + \alpha_{11}^{(2)} x_1^2 = 0. \tag{10}$$

We note that Eqs. (9)-(10) are the normal form of the conservative dynamical system of interest, with $\gamma_{111}^{(1)} = \gamma$ being the Duffing nonlinearity of the first mode, and $\alpha = \alpha_{12}^{(1)}/2 = \alpha_{11}^{(2)}$ is the coupling coefficient, which stems from a single-term potential $U_{cp} = \alpha x_1^2 x_2$.

## SUPPLEMENTAL MATERIAL III. SLOW DYNAMIC EQUATIONS AND BIFURCATION ANALYSIS

In order to investigate the experimentally observed physics in our graphene nanodrum, we use the normal form of the equations obtained in the previous section in the presence of damping and external forcing terms. The resulting equations then read

$$\ddot{x}_1 + x_1 + \gamma x_1^3 + 2\alpha x_2 x_1 = F_1 x_1 \cos(\omega_F t) - 2\tau_1 \dot{x}_1 - 2\tau_{nl1} x_1^2 \dot{x}_1, \tag{11}$$

$$\ddot{x}_2 + \omega_2^2 x_2 + \alpha x_1^2 = F_2 \cos(\omega_F t) - 2\tau_2 \dot{x}_2, \tag{12}$$

in which $\tau_1$ and $\tau_2$ are the damping coefficients associated with modes 1 and 2 , respectively. $F_1$ is the parametric drive, $F_2$ is the direct drive, and $\tau_{nl1}$ is the van der Pol type nonlinear damping term added to the equation of motion to avoid unbounded parametric resonance [4].

Applying the rotating wave approximation (RWA) $x_1(t) = A_1(t)\exp(i\omega_F t/2) + A_1^*(t)\exp(-i\omega_F t/2)$, $\dot{x}_1(t) = (i\omega_F/2)[A_1(t)\exp(i\omega_F t/2) - A_1^*(t)\exp(-i\omega_F t/2)]$, $x_2 = A_2(t)\exp(i\omega_F t) + A_2^*(t)\exp(-i\omega_F t)$, $\dot{x}_2 = i\omega_F[A_2(t)\exp(i\omega_F t) - A_2^*(t)\exp(-i\omega_F t)]$, where $A_j$ and $A_j^*$ are the complex-amplitude of the $j^{th}$ mode and its complex-conjugate, respectively, and introducing the detuning parameters $\Delta\omega_1 = \omega_F/2 - \omega_1$, $\Delta\omega_2 = \omega_F - \omega_2$, we obtain the following evolution equations for the complex amplitudes

$$\dot{A}_1 = -\left[\tau_1 + \tau_{nl1}|A_1|^2 + i\left(\Delta\omega_1 - \frac{3\gamma}{\omega_F}|A_1|^2\right)\right]A_1 + \frac{2i}{\omega_F}\left(\alpha A_2 - \frac{F_1}{4}\right)A_1^*, \tag{13}$$

$$\dot{A}_2 = -(\tau_2 + i\Delta\omega_2)A_2 + \frac{i}{2\omega_F}\left(\alpha A_1^2 - \frac{F_2}{2}\right). \tag{14}$$

At steady-state $\dot{A}_j = 0$. Thus, from Eq. (14) we find that the steady-state complex-amplitude of the second mode $A_{2ss}$ is given by

$$A_{2ss} = \frac{(i\tau_2 + \Delta\omega_2)}{2\omega_F(\tau_2^2 + \Delta\omega_2^2)}\left(\alpha A_{1ss}^2 - \frac{F_2}{2}\right). \tag{15}$$

Hence, by substituting Eq. (15) into Eq. (13), we obtain a single equation for the steady-state complex-amplitude of the first mode $A_{1ss}$ as follows

$$\left[\tau_1 + \left(\tau_{nl1} + \frac{\alpha^2\tau_2}{\omega_F^2(\tau_2^2 + \Delta\omega_2^2)}\right)|A_{1ss}|^2\right]A_{1ss} + i\left[\Delta\omega_1 - \left(\frac{3\gamma}{\omega_F} + \frac{\alpha^2\Delta\omega_2}{\omega_F^2(\tau_2^2 + \Delta\omega_2^2)}\right)|A_{1ss}|^2\right]A_{1ss}$$
$$= -\frac{i}{2\omega_F}\left(F_1 + \frac{\alpha(i\tau_2 + \Delta\omega_2)}{\omega_F(\tau_2^2 + \Delta\omega_2^2)}F_2\right)A_{1ss}^*. \tag{16}$$

Taking the magnitude squared of both sides of Eq. (16), we find the intensity of the non-trivial response of the first mode $4|A_{1ss}|^2 \equiv a_1^2$ as the positive roots of the equation

$$\left[\tau_1 + \left(\tau_{nl1} + \frac{\alpha^2\tau_2}{\omega_F^2(\tau_2^2 + \Delta\omega_2^2)}\right)\frac{a_1^2}{4}\right]^2 + \left[\Delta\omega_1 - \left(\frac{3\gamma}{\omega_F} + \frac{\alpha^2\Delta\omega_2}{\omega_F^2(\tau_2^2 + \Delta\omega_2^2)}\right)\frac{a_1^2}{4}\right]^2$$
$$= \frac{1}{4\omega_F^2}\left[F_1^2 + \frac{\alpha F_2(\alpha F_2 + 2\omega_F\Delta\omega_2 F_1)}{\omega_F^2(\tau_2^2 + \Delta\omega_2^2)}\right]. \tag{17}$$

We note that Eq. (17) is a quadratic equation in $a_1^2$. Thus, in addition to the trivial solution $a_1 = 0$, which is always a solution, we are left with, at most, two additional positive solutions for $a_1$. Therefore, the non-trivial steady-state solutions of the first mode $x_1 = a_1 \cos(\omega_F t/2 + \phi_1)$ is fully described by Eq. (17) and the following equation for the phase

$$e^{2i\phi_1} = \frac{A_{1ss}}{A_{1ss}^*} = \frac{-\frac{i}{2\omega_F}\left(F_1 + \frac{\alpha(i\tau_2 + \Delta\omega_2)}{\omega_F(\tau_2^2 + \Delta\omega_2^2)}F_2\right)}{\tau_1 + \left(\tau_{nl1} + \frac{\alpha^2\tau_2}{\omega_F^2(\tau_2^2 + \Delta\omega_2^2)}\right)\frac{a_1^2}{4} + i\left[\Delta\omega_1 - \left(\frac{3\gamma}{\omega_F} + \frac{\alpha^2\Delta\omega_2}{\omega_F^2(\tau_2^2 + \Delta\omega_2^2)}\right)\frac{a_1^2}{4}\right]}. \tag{18}$$

From Eq. (15) we also find the steady-state solution of the second mode $A_{2ss} = a_2 e^{i\phi_2}/2$. Thus, the solution $a_1$ of Eq. (17) along with Eqs. (15), (18) give complete description of the system non-trivial steady-state solutions. Note

that the peak amplitude of the first mode $A_{SNB}$ can be calculated by differentiation of Eq. (17) with respect to $\omega_F$ and setting $da_1^2/d\omega_F = 0$. To the leading order approximation, the expression for the peak amplitude $A_{SNB}$ is given by

$$A_{\mathrm{SNB}}^2 = \frac{2\sqrt{\tau_2^2 + \Delta\omega_2^2}\left[\omega_F^2(\tau_2^2 + \Delta\omega_2^2)F_1^2 + \alpha F_2(\alpha F_2 + 2\omega_F\Delta\omega_2 F_1)\right]^{1/2} - 4\omega_F^2(\tau_2^2 + \Delta\omega_2^2)\tau_1}{\omega_F^2(\tau_2^2 + \Delta\omega_2^2)\tau_{nl1} + \alpha^2\tau_2}. \tag{19}$$

### Instability threshold of the trivial solution

In order to find the conditions under which the trivial solution of the first mode $a_1 = 0$ loses its stability, we linearize Eqs. (13)-(14) around $A_{1ss} = 0$, $A_{2ss} = -\frac{(i\tau_2 + \Delta\omega_2)F_2}{4\omega_F(\tau_2^2 + \Delta\omega_2^2)}$, and obtain the eigenvalues of the resulting linear system

$$\delta\dot{A}_1 = -\left[\tau_1 + i\Delta\omega_1\right]\delta A_1 - \frac{i}{2\omega_F}\left(F_1 + \frac{\alpha(i\tau_2 + \Delta\omega_2)}{\omega_F(\tau_2^2 + \Delta\omega_2^2)}F_2\right)\delta\bar{A}_1, \tag{20}$$

$$\delta\dot{A}_2 = -(\tau_2 + i\Delta\omega_2)\delta A_2. \tag{21}$$

Note that Eqs. (20)-(21) are uncoupled, and hence, we can analyze each equation separately. Eq. (21) can be readily solved to yield $\delta A_2 = \delta A_2(0)e^{-(\tau_2 + i\Delta\omega_2)t}$. Consequently, $\delta A_2$ decays to zero for all $\tau_2 > 0$. Similarly, we seek a solution for Eq. (20) in the form $\delta A_1 = |\delta A_1|e^{i\delta\phi_1}e^{\lambda t}$, where $\delta\phi_1$ and $\lambda$ are real. By substitution of the solution into Eq. (20) we find that

$$\lambda = -\tau_1 \pm \sqrt{\frac{1}{4\omega_F^2}\left[F_1^2 + \frac{\alpha F_2(\alpha F_2 + 2\omega_F\Delta\omega_2 F_1)}{\omega_F^2(\tau_2^2 + \Delta\omega_2^2)}\right] - \Delta\omega_1^2}. \tag{22}$$

Thus, for $\lambda > 0$, the trivial solution of the first mode $a_1 = 0$ is unstable (i.e., there is a supercritical pitchfork bifurcation). The boundaries, which define the domains of the instability, form the Mathieu tongue and are given by

$$4\omega_F^2(\tau_1^2 + \Delta\omega_1^2) = F_1^2 + \frac{\alpha F_2(\alpha F_2 + 2\omega_F\Delta\omega_2 F_1)}{\omega_F^2(\tau_2^2 + \Delta\omega_2^2)}. \tag{23}$$

Note that the same condition can be obtained from Eq. (17) by taking the limit $a_1 \to 0$.

### Stability of the non-trivial steady-state solutions and local bifurcation analysis

To investigate the stability of the non-trivial steady-state solutions [which are found from Eq. (15) and Eqs. (17)-(18)], we superimpose a perturbation $\delta\mathbf{u} = (\delta A_1, \delta A_2)^T$ on the non-trivial fixed-points of Eq. (13)-(14) $\mathbf{u}_{ss} = (A_{1ss}, A_{2ss})^T$, linearize in terms of the perturbed variables, and obtain the following pair of linear complex evolution-equations for the perturbation $\delta\mathbf{u}$

$$\delta\dot{A}_1 = -\left[\tau_1 + 2\tau_{nl1}|A_{1ss}|^2 + i\left(\Delta\omega_1 - \frac{6\gamma}{\omega_F}|A_{1ss}|^2\right)\right]\delta A_1$$
$$- \left[\tau_{nl1}A_{1ss}^2 + \frac{i}{\omega_F}\left(\frac{F_1}{2} - 2\alpha A_{2ss} - 3\gamma A_{1ss}^2\right)\right]\delta\bar{A}_1 + \frac{2i\alpha}{\omega_F}\bar{A}_{1ss}\delta A_2, \tag{24}$$

$$\delta\dot{A}_2 = \frac{i\alpha}{\omega_F}A_{1ss}\delta A_1 - (\tau_2 + i\Delta\omega_2)\delta A_2. \tag{25}$$

Using Cartesian notations for the perturbations $\delta A_j = q_j + ip_j$, Eqs. (24)-(25) can be written as $\dot{q}_1 + i\dot{p}_1 = f_1 + if_2$ and $\dot{q}_2 + i\dot{p}_2 = f_3 + if_4$. Thus, by taking the real and imaginary part of these equations, we obtain a set of four linear real evolution-equations for the perturbation quadratures $\dot{\boldsymbol{\eta}} = \mathbf{J} \cdot \boldsymbol{\eta}$, where $\boldsymbol{\eta} = (q_1, p_1, q_2, p_2)^T$, $J_{nm} = \partial f_n/\partial\eta_m|_{A_{1ss}, A_{2ss}}$. Therefore, the stability of the fixed points are determined by the eigenvalues $\lambda$ of $\mathbf{J}$, which are the roots of the following characteristic polynomial

$$\lambda^4 + c_1\lambda^3 + c_2\lambda^2 + c_3\lambda + c_4 = 0$$
$$c_1 = -\mathrm{tr}(\mathbf{J}), \;\; c_2 = \frac{1}{2}(\mathrm{tr}(\mathbf{J})^2 - \mathrm{tr}(\mathbf{J}^2)), \;\; c_3 = -\frac{1}{3}\left(\mathrm{tr}(\mathbf{J}^3) - \frac{3}{2}\mathrm{tr}(\mathbf{J}^2)\mathrm{tr}(\mathbf{J}) + \mathrm{tr}(\mathbf{J})^3\right), \;\; c_4 = \det(\mathbf{J}). \tag{26}$$

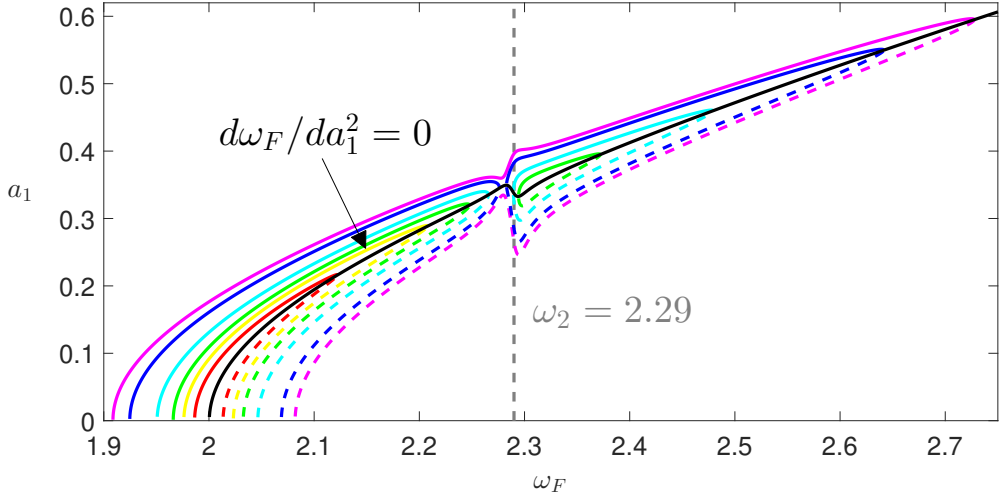

FIG. 2: Response curve of the first mode for $\tau_1 = 0.0025, \tau_{nl1} = 0.175, \tau_2 = 0.00764, \alpha = 0.17, \gamma = 3.7, \omega_1 = 1, \omega_2 = 2.29, F_2 = 0.13F_1$. The magenta/dark-blue/light-blue/green/yellow/red curves are the first mode response curve for drive amplitudes $F_1 = 0.18, 0.15, 0.1, 0.07, 0.05, 0.03$. The grey vertical dashed line shows the frequency of the second mode, and the black curve show the loci of the saddle-node bifurcations.

Consequently, there is a saddle-node bifurcation whenever $c_4 = 0$, and Hopf bifurcation whenever $\Delta_3 = c_3(c_2c_1 - c_3) - c_4c_1^2 = 0$. Furthermore, near the Hopf threshold ($\Delta_3 = 0$), the frequency of the limit cycle is given by $\lambda_{1,2} = \pm i\Omega_H$, $\Omega_H^2 = c_3/c_1$. Note that the condition for the saddle-node bifurcation $c_4 = 0$ can also be obtained by differentiating Eq. (17) with respect to the drive frequency $\omega_F$, and requiring that $d\omega_F/da_1^2 = 0$ (i.e., requirement of vertical tangency in the response curve). Fig. 2 shows the response curves for several values of drive amplitude $F_1$ along with the locus of the saddle-node bifurcations, and an indication of the second mode frequency (assuming that $\omega_2 = 2.29\omega_1$). It is clear from the figure that as the driving amplitude $F_1$ increases, the gap between the first two saddle-node bifurcations decreases. At a critical drive level (the magenta response curve), these two saddle-node bifurcations annihilate each other and the response becomes continuous.

## SUPPLEMENTAL MATERIAL IV. MODEL FITS USING THE SINGLE-MODE EQUATION

In this section, we use Eq. (11) in the absence of quadratic coupling ($\alpha = 0$) to fit the experimental data. The fittings are carried out by using a sequential algorithm based on pseudo arc-length continuation and numerical integration [5]. To fit the data, the information about resonance frequency $\omega_1$ and damping coefficient $\tau_1$ is directly extracted from the measurements. The linear quality factor Q of the resonator is determined at low drive level to be 454 and has been fixed throughout the fitting procedure. Moreover, the magnitude of the parametric drive $F_1$ at low driving powers is estimated by tracing the locus of the pitch-fork bifurcation points (Mathieu tongue) as shown in Fig. 3. However, for higher power levels, the bifurcation points did not give reliable information due to frequency fluctuations. Thus, to obtain the parametric forces at high drive powers, we use lower force levels (known from Mathieu tongue) and extrapolate using the ratios between the applied voltage/power levels from the VNA and the forces such that $F_{1,\text{high}} = F_{1,\text{low}}\frac{P_{\text{high}}}{P_{\text{low}}}$. In other words, we assume that the drive levels applied in the experiments using the VNA are linearly related to the force felt by the resonator.

Once $F_1$ is estimated, $\gamma$ is fitted by minimizing the curvatures of the hardening type nonlinearity observed in the experiments and the model. Finally, to match the peak amplitudes observed in the experiments, the nonlinear damping coefficient $\tau_{nl1}$ is used as the fitting parameter such that the saddle-node bifurcation amplitude ($A_{\text{SNB}}$) and frequency ($f_{\text{SNB}}$) are within 0.1% error of the $A_{\text{SNB}}$ and $f_{\text{SNB}}$ obtained experimentally. In Fig. 4 we report the fitted curves and their associated $\gamma$ and $\tau_{nl1}$.

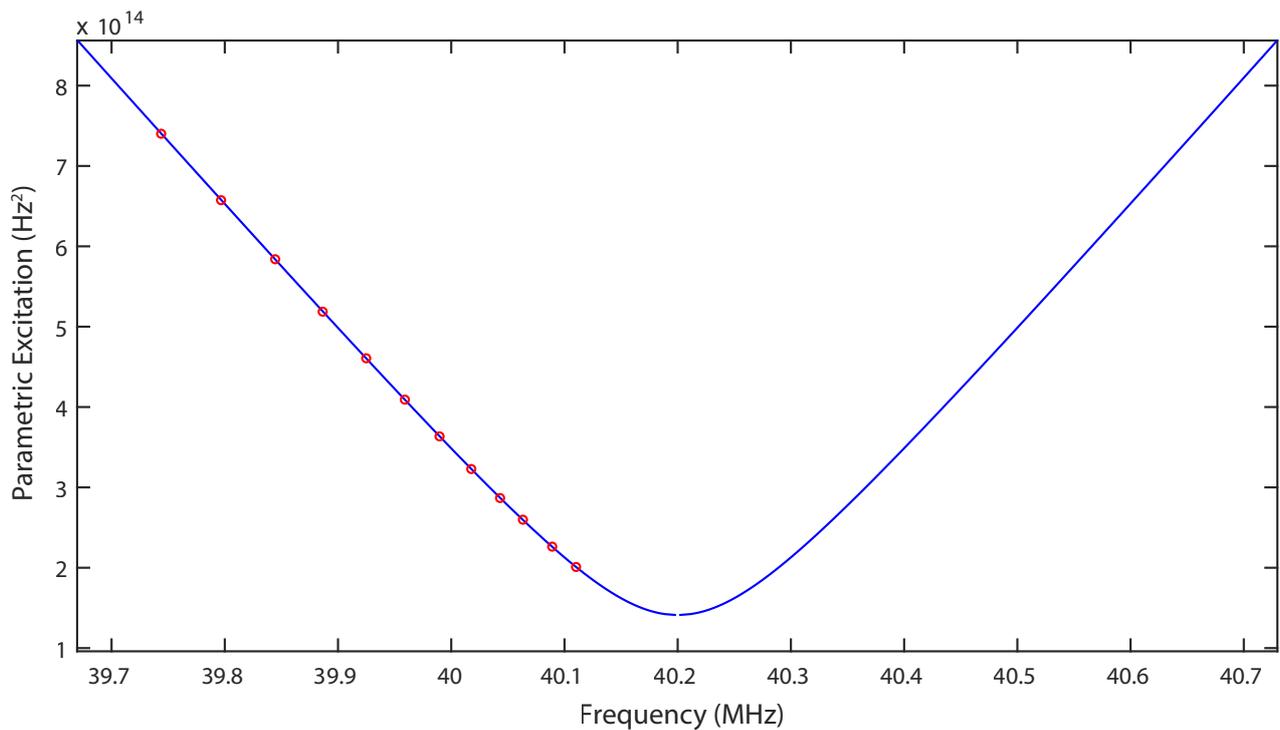

FIG. 3: The Mathieu tongue associated with the experimental pitch-fork bifurcation frequencies

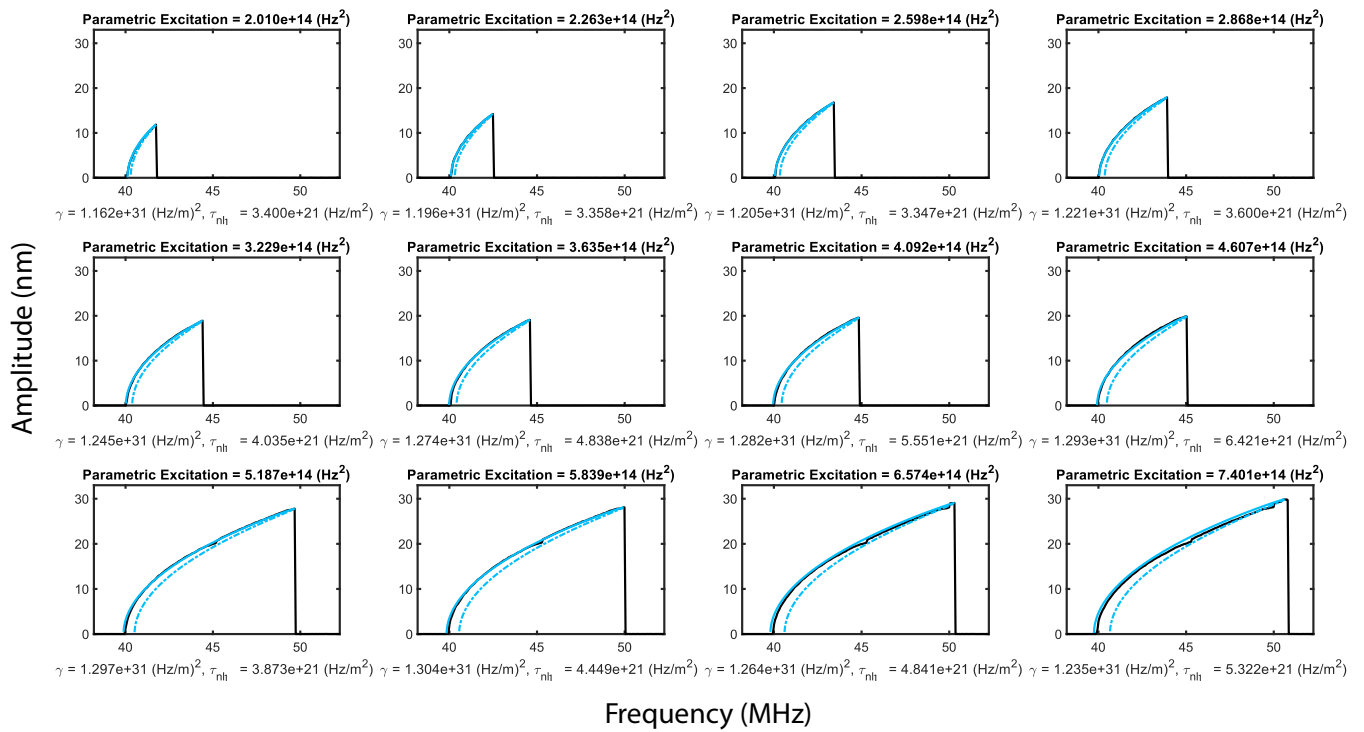

FIG. 4: Fits obtained using the single-mode model

## SUPPLEMENTAL MATERIAL V. MODEL FITS USING THE COUPLED MODEL

To fit with the coupled model, Eq. (11) and (12) are used. The parameters from the single-mode fits in the uncoupled regime (at the force levels where the coupling effects are negligible) have been taken as base parameters, since they are intrinsic to the modes themselves. This leaves only 2 additional parameters for fitting, namely the coupling strength $\alpha$ and the direct forcing $F_2$. Before the interaction, $F_2$ can be directly obtained from the measurements, however during the interaction the individual amplitude of the second mode is hidden beneath the parametric resonance curve. Thus, we assume a linear relation between $F_1$ and $F_2$ such that $F_2 = c_{pm}F_1$ so as to estimate the contribution of $F_2$ in the coupled mode dynamics. Finally, we use $\alpha$ as the fit parameter and minimize the error between the resonance peak ($A_{\text{SNB}}$) of the model and experimental data.

In Fig. 5 we report the fits to the experimental data using the coupled model , and in Table 1, we report the fitted values. We note that the curves with different parametric excitation are fitted with constant parameters.

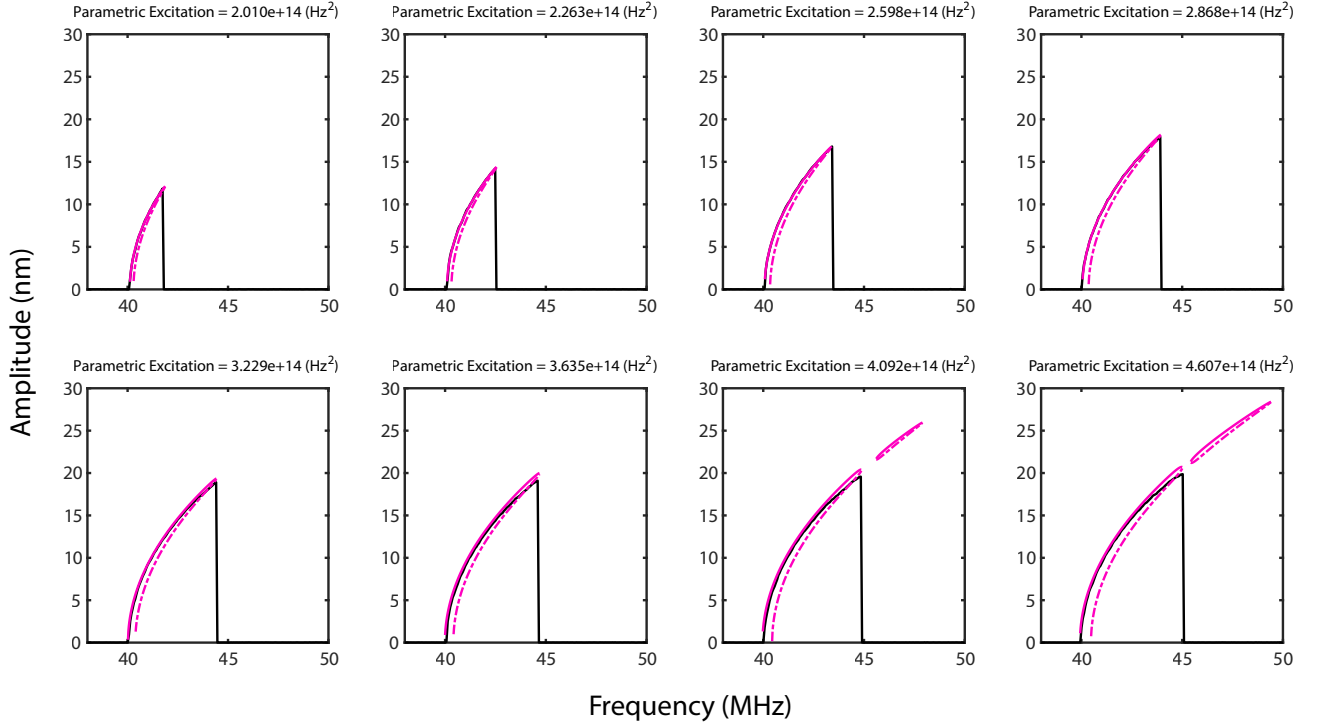

FIG. 5: Fits obtained using the 2 DoF coupled model

TABLE I: Parameter set extracted from the fits of Fig. 5

| $w_1$ | $w_2$ | $\tau_1$ | $\tau_2$ | $\tau_{nl1}$ | $\gamma$ | $\alpha$ | $c_{pm}$ |
|---|---|---|---|---|---|---|---|
| $1.262 \times 10^8$ (rad/s) | $2.841 \times 10^8$ (rad/s) | $2.781 \times 10^5$ (Hz) | $2.185 \times 10^6$ (Hz) | $3.139 \times 10^{21}$ (Hz/m²) | $1.205 \times 10^{31}$ (Hz/m)² | $2.213 \times 10^{22}$ (Hz²/m) | $1.585 \times 10^{-9}$ (m) |

---

[1] R. J. Dolleman, D. Davidovikj, H. S. J. van der Zant, and P. G. Steeneken, "Amplitude calibration of 2d mechanical resonators by nonlinear optical transduction," *Applied Physics Letters*, vol. 111, no. 25, p. 253104, 2017.

[2] M. Amabili, *Nonlinear vibrations and stability of shells and plates.* Cambridge University Press, 2008.

[3] J. Guckenheimer and P. Holmes, *Nonlinear Oscillations, Dynamical Systems, and Bifurcations of Vector Fields.* Applied Mathematical Sciences, Springer New York, 2013.

[4] R. Lifshitz and M. Cross, "Nonlinear dynamics of nanomechanical and micromechanical resonators," *Review of nonlinear dynamics and complexity*, vol. 1, pp. 1–52, 2008.

[5] E. J. Doedel, A. R. Champneys, T. F. Fairgrieve, Y. A. Kuznetsov, B. Sandstede, X. Wang, *et al.*, "Auto97," *Continuation and bifurcation software for ordinary differential equations*, 1998.

\AtBeginShipout\AtBeginShipoutDiscard


\begin{thebibliography}{33}
	\expandafter\ifx\csname natexlab\endcsname\relax\def\natexlab#1{#1}\fi
	\expandafter\ifx\csname bibnamefont\endcsname\relax
	\def\bibnamefont#1{#1}\fi
	\expandafter\ifx\csname bibfnamefont\endcsname\relax
	\def\bibfnamefont#1{#1}\fi
	\expandafter\ifx\csname citenamefont\endcsname\relax
	\def\citenamefont#1{#1}\fi
	\expandafter\ifx\csname url\endcsname\relax
	\def\url#1{\texttt{#1}}\fi
	\expandafter\ifx\csname urlprefix\endcsname\relax\def\urlprefix{URL }\fi
	\providecommand{\bibinfo}[2]{#2}
	\providecommand{\eprint}[2][]{\url{#2}}
	
	\bibitem[{\citenamefont{Eichler et~al.}(2011)\citenamefont{Eichler, Moser,
			Chaste, Zdrojek, Wilson-Rae, and Bachtold}}]{bachtoldfirst}
	\bibinfo{author}{\bibfnamefont{A.}~\bibnamefont{Eichler}},
	\bibinfo{author}{\bibfnamefont{J.}~\bibnamefont{Moser}},
	\bibinfo{author}{\bibfnamefont{J.}~\bibnamefont{Chaste}},
	\bibinfo{author}{\bibfnamefont{M.}~\bibnamefont{Zdrojek}},
	\bibinfo{author}{\bibfnamefont{I.}~\bibnamefont{Wilson-Rae}},
	\bibnamefont{and} \bibinfo{author}{\bibfnamefont{A.}~\bibnamefont{Bachtold}},
	\bibinfo{journal}{Nature Nanotechnology} \textbf{\bibinfo{volume}{6}},
	\bibinfo{pages}{339} (\bibinfo{year}{2011}).
	
	\bibitem[{\citenamefont{Amabili}(2019)}]{Amabili2019}
	\bibinfo{author}{\bibfnamefont{M.}~\bibnamefont{Amabili}},
	\bibinfo{journal}{Nonlinear Dynamics} \textbf{\bibinfo{volume}{97}},
	\bibinfo{pages}{1785} (\bibinfo{year}{2019}).
	
	\bibitem[{\citenamefont{Amabili et~al.}(2020)\citenamefont{Amabili,
			Balasubramanian, Bozzo, Breslavsky, Ferrari, Franchini, Giovanniello, and
			Pogue}}]{amabili2020nonlinear}
	\bibinfo{author}{\bibfnamefont{M.}~\bibnamefont{Amabili}},
	\bibinfo{author}{\bibfnamefont{P.}~\bibnamefont{Balasubramanian}},
	\bibinfo{author}{\bibfnamefont{I.}~\bibnamefont{Bozzo}},
	\bibinfo{author}{\bibfnamefont{I.~D.} \bibnamefont{Breslavsky}},
	\bibinfo{author}{\bibfnamefont{G.}~\bibnamefont{Ferrari}},
	\bibinfo{author}{\bibfnamefont{G.}~\bibnamefont{Franchini}},
	\bibinfo{author}{\bibfnamefont{F.}~\bibnamefont{Giovanniello}},
	\bibnamefont{and} \bibinfo{author}{\bibfnamefont{C.}~\bibnamefont{Pogue}},
	\bibinfo{journal}{Physical Review X} \textbf{\bibinfo{volume}{10}},
	\bibinfo{pages}{011015} (\bibinfo{year}{2020}).
	
	\bibitem[{\citenamefont{Midtvedt et~al.}(2014)\citenamefont{Midtvedt, Croy,
			Isacsson, Qi, and Park}}]{midtvedt2014fermi}
	\bibinfo{author}{\bibfnamefont{D.}~\bibnamefont{Midtvedt}},
	\bibinfo{author}{\bibfnamefont{A.}~\bibnamefont{Croy}},
	\bibinfo{author}{\bibfnamefont{A.}~\bibnamefont{Isacsson}},
	\bibinfo{author}{\bibfnamefont{Z.}~\bibnamefont{Qi}}, \bibnamefont{and}
	\bibinfo{author}{\bibfnamefont{H.~S.} \bibnamefont{Park}},
	\bibinfo{journal}{Physical review letters} \textbf{\bibinfo{volume}{112}},
	\bibinfo{pages}{145503} (\bibinfo{year}{2014}).
	
	\bibitem[{\citenamefont{Leghtas et~al.}(2015)\citenamefont{Leghtas, Touzard,
			Pop, Kou, Vlastakis, Petrenko, Sliwa, Narla, Shankar, Hatridge
			et~al.}}]{leghtas2015confining}
	\bibinfo{author}{\bibfnamefont{Z.}~\bibnamefont{Leghtas}},
	\bibinfo{author}{\bibfnamefont{S.}~\bibnamefont{Touzard}},
	\bibinfo{author}{\bibfnamefont{I.~M.} \bibnamefont{Pop}},
	\bibinfo{author}{\bibfnamefont{A.}~\bibnamefont{Kou}},
	\bibinfo{author}{\bibfnamefont{B.}~\bibnamefont{Vlastakis}},
	\bibinfo{author}{\bibfnamefont{A.}~\bibnamefont{Petrenko}},
	\bibinfo{author}{\bibfnamefont{K.~M.} \bibnamefont{Sliwa}},
	\bibinfo{author}{\bibfnamefont{A.}~\bibnamefont{Narla}},
	\bibinfo{author}{\bibfnamefont{S.}~\bibnamefont{Shankar}},
	\bibinfo{author}{\bibfnamefont{M.~J.} \bibnamefont{Hatridge}},
	\bibnamefont{et~al.}, \bibinfo{journal}{Science}
	\textbf{\bibinfo{volume}{347}}, \bibinfo{pages}{853} (\bibinfo{year}{2015}).
	
	\bibitem[{\citenamefont{Zaitsev et~al.}(2012)\citenamefont{Zaitsev, Shtempluck,
			Buks, and Gottlieb}}]{viscoZaitsev}
	\bibinfo{author}{\bibfnamefont{S.}~\bibnamefont{Zaitsev}},
	\bibinfo{author}{\bibfnamefont{O.}~\bibnamefont{Shtempluck}},
	\bibinfo{author}{\bibfnamefont{E.}~\bibnamefont{Buks}}, \bibnamefont{and}
	\bibinfo{author}{\bibfnamefont{O.}~\bibnamefont{Gottlieb}},
	\bibinfo{journal}{Nonlinear Dynamics} \textbf{\bibinfo{volume}{67}},
	\bibinfo{pages}{859} (\bibinfo{year}{2012}).
	
	\bibitem[{\citenamefont{Croy et~al.}(2012)\citenamefont{Croy, Midtvedt,
			Isacsson, and Kinaret}}]{croyPhonon}
	\bibinfo{author}{\bibfnamefont{A.}~\bibnamefont{Croy}},
	\bibinfo{author}{\bibfnamefont{D.}~\bibnamefont{Midtvedt}},
	\bibinfo{author}{\bibfnamefont{A.}~\bibnamefont{Isacsson}}, \bibnamefont{and}
	\bibinfo{author}{\bibfnamefont{J.~M.} \bibnamefont{Kinaret}},
	\bibinfo{journal}{Phys. Rev. B} \textbf{\bibinfo{volume}{86}},
	\bibinfo{pages}{235435} (\bibinfo{year}{2012}).
	
	\bibitem[{\citenamefont{Atalaya et~al.}(2016)\citenamefont{Atalaya, Kenny,
			Roukes, and Dykman}}]{atalaya2016nonlinear}
	\bibinfo{author}{\bibfnamefont{J.}~\bibnamefont{Atalaya}},
	\bibinfo{author}{\bibfnamefont{T.~W.} \bibnamefont{Kenny}},
	\bibinfo{author}{\bibfnamefont{M.}~\bibnamefont{Roukes}}, \bibnamefont{and}
	\bibinfo{author}{\bibfnamefont{M.}~\bibnamefont{Dykman}},
	\bibinfo{journal}{Physical Review B} \textbf{\bibinfo{volume}{94}},
	\bibinfo{pages}{195440} (\bibinfo{year}{2016}).
	
	\bibitem[{\citenamefont{G{\"u}ttinger et~al.}(2017)\citenamefont{G{\"u}ttinger,
			Noury, Weber, Eriksson, Lagoin, Moser, Eichler, Wallraff, Isacsson, and
			Bachtold}}]{guttinger2017energy}
	\bibinfo{author}{\bibfnamefont{J.}~\bibnamefont{G{\"u}ttinger}},
	\bibinfo{author}{\bibfnamefont{A.}~\bibnamefont{Noury}},
	\bibinfo{author}{\bibfnamefont{P.}~\bibnamefont{Weber}},
	\bibinfo{author}{\bibfnamefont{A.~M.} \bibnamefont{Eriksson}},
	\bibinfo{author}{\bibfnamefont{C.}~\bibnamefont{Lagoin}},
	\bibinfo{author}{\bibfnamefont{J.}~\bibnamefont{Moser}},
	\bibinfo{author}{\bibfnamefont{C.}~\bibnamefont{Eichler}},
	\bibinfo{author}{\bibfnamefont{A.}~\bibnamefont{Wallraff}},
	\bibinfo{author}{\bibfnamefont{A.}~\bibnamefont{Isacsson}}, \bibnamefont{and}
	\bibinfo{author}{\bibfnamefont{A.}~\bibnamefont{Bachtold}},
	\bibinfo{journal}{Nature nanotechnology} \textbf{\bibinfo{volume}{12}},
	\bibinfo{pages}{631} (\bibinfo{year}{2017}).
	
	\bibitem[{\citenamefont{Song et~al.}(2012)\citenamefont{Song, Oksanen,
			Sillanpää, Craighead, Parpia, and Hakonen}}]{song2012stamp}
	\bibinfo{author}{\bibfnamefont{X.}~\bibnamefont{Song}},
	\bibinfo{author}{\bibfnamefont{M.}~\bibnamefont{Oksanen}},
	\bibinfo{author}{\bibfnamefont{M.~A.} \bibnamefont{Sillanpää}},
	\bibinfo{author}{\bibfnamefont{H.}~\bibnamefont{Craighead}},
	\bibinfo{author}{\bibfnamefont{J.}~\bibnamefont{Parpia}}, \bibnamefont{and}
	\bibinfo{author}{\bibfnamefont{P.~J.} \bibnamefont{Hakonen}},
	\bibinfo{journal}{Nano letters} \textbf{\bibinfo{volume}{12}},
	\bibinfo{pages}{198} (\bibinfo{year}{2012}).
	
	\bibitem[{\citenamefont{Sajadi et~al.}(2017)\citenamefont{Sajadi, Alijani,
			Davidovikj, Goosen, Steeneken, and van Keulen}}]{sajadi2017experimental}
	\bibinfo{author}{\bibfnamefont{B.}~\bibnamefont{Sajadi}},
	\bibinfo{author}{\bibfnamefont{F.}~\bibnamefont{Alijani}},
	\bibinfo{author}{\bibfnamefont{D.}~\bibnamefont{Davidovikj}},
	\bibinfo{author}{\bibfnamefont{J.}~\bibnamefont{Goosen}},
	\bibinfo{author}{\bibfnamefont{P.~G.} \bibnamefont{Steeneken}},
	\bibnamefont{and} \bibinfo{author}{\bibfnamefont{F.}~\bibnamefont{van
			Keulen}}, \bibinfo{journal}{Journal of Applied Physics}
	\textbf{\bibinfo{volume}{122}}, \bibinfo{pages}{234302}
	(\bibinfo{year}{2017}).
	
	\bibitem[{\citenamefont{Lee et~al.}(2019)\citenamefont{Lee, Davidovikj, Sajadi,
			\v{S}i\v{s}kins, Alijani, van~der Zant, and Steeneken}}]{lee2019sealing}
	\bibinfo{author}{\bibfnamefont{M.}~\bibnamefont{Lee}},
	\bibinfo{author}{\bibfnamefont{D.}~\bibnamefont{Davidovikj}},
	\bibinfo{author}{\bibfnamefont{B.}~\bibnamefont{Sajadi}},
	\bibinfo{author}{\bibfnamefont{M.}~\bibnamefont{\v{S}i\v{s}kins}},
	\bibinfo{author}{\bibfnamefont{F.}~\bibnamefont{Alijani}},
	\bibinfo{author}{\bibfnamefont{H.~S.} \bibnamefont{van~der Zant}},
	\bibnamefont{and} \bibinfo{author}{\bibfnamefont{P.~G.}
		\bibnamefont{Steeneken}}, \bibinfo{journal}{Nano letters}
	\textbf{\bibinfo{volume}{19}}, \bibinfo{pages}{5313} (\bibinfo{year}{2019}).
	
	\bibitem[{\citenamefont{Miller et~al.}(2018)\citenamefont{Miller, Ansari,
			Heinz, Chen, Flader, Shin, Villanueva, and Kenny}}]{miller2018effective}
	\bibinfo{author}{\bibfnamefont{J.~M.~L.} \bibnamefont{Miller}},
	\bibinfo{author}{\bibfnamefont{A.}~\bibnamefont{Ansari}},
	\bibinfo{author}{\bibfnamefont{D.~B.} \bibnamefont{Heinz}},
	\bibinfo{author}{\bibfnamefont{Y.}~\bibnamefont{Chen}},
	\bibinfo{author}{\bibfnamefont{I.~B.} \bibnamefont{Flader}},
	\bibinfo{author}{\bibfnamefont{D.~D.} \bibnamefont{Shin}},
	\bibinfo{author}{\bibfnamefont{L.~G.} \bibnamefont{Villanueva}},
	\bibnamefont{and} \bibinfo{author}{\bibfnamefont{T.~W.} \bibnamefont{Kenny}},
	\bibinfo{journal}{Applied Physics Reviews} \textbf{\bibinfo{volume}{5}},
	\bibinfo{pages}{041307} (\bibinfo{year}{2018}).
	
	\bibitem[{\citenamefont{Weber et~al.}(2014)\citenamefont{Weber, Guttinger,
			Tsioutsios, Chang, and Bachtold}}]{weber2014coupling}
	\bibinfo{author}{\bibfnamefont{P.}~\bibnamefont{Weber}},
	\bibinfo{author}{\bibfnamefont{J.}~\bibnamefont{Guttinger}},
	\bibinfo{author}{\bibfnamefont{I.}~\bibnamefont{Tsioutsios}},
	\bibinfo{author}{\bibfnamefont{D.~E.} \bibnamefont{Chang}}, \bibnamefont{and}
	\bibinfo{author}{\bibfnamefont{A.}~\bibnamefont{Bachtold}},
	\bibinfo{journal}{Nano letters} \textbf{\bibinfo{volume}{14}},
	\bibinfo{pages}{2854} (\bibinfo{year}{2014}).
	
	\bibitem[{\citenamefont{Samanta et~al.}(2018)\citenamefont{Samanta, Arora, and
			Naik}}]{samanta2018tuning}
	\bibinfo{author}{\bibfnamefont{C.}~\bibnamefont{Samanta}},
	\bibinfo{author}{\bibfnamefont{N.}~\bibnamefont{Arora}}, \bibnamefont{and}
	\bibinfo{author}{\bibfnamefont{A.}~\bibnamefont{Naik}},
	\bibinfo{journal}{Applied Physics Letters} \textbf{\bibinfo{volume}{113}},
	\bibinfo{pages}{113101} (\bibinfo{year}{2018}).
	
	\bibitem[{\citenamefont{Yang et~al.}(2020)\citenamefont{Yang, Rochau, Huber,
			Brieussel, Rastelli, Weig, and Scheer}}]{yang2020persistent}
	\bibinfo{author}{\bibfnamefont{F.}~\bibnamefont{Yang}},
	\bibinfo{author}{\bibfnamefont{F.}~\bibnamefont{Rochau}},
	\bibinfo{author}{\bibfnamefont{J.~S.} \bibnamefont{Huber}},
	\bibinfo{author}{\bibfnamefont{A.}~\bibnamefont{Brieussel}},
	\bibinfo{author}{\bibfnamefont{G.}~\bibnamefont{Rastelli}},
	\bibinfo{author}{\bibfnamefont{E.~M.} \bibnamefont{Weig}}, \bibnamefont{and}
	\bibinfo{author}{\bibfnamefont{E.}~\bibnamefont{Scheer}},
	\bibinfo{journal}{arXiv preprint arXiv:2003.14207}  (\bibinfo{year}{2020}).
	
	\bibitem[{\citenamefont{Nayfeh and Mook}(1995)}]{nayfnonosc}
	\bibinfo{author}{\bibfnamefont{A.~H.} \bibnamefont{Nayfeh}} \bibnamefont{and}
	\bibinfo{author}{\bibfnamefont{D.~T.} \bibnamefont{Mook}},
	\emph{\bibinfo{title}{Nonlinear Oscillations}} (\bibinfo{publisher}{John
		Wiley \& Sons}, \bibinfo{year}{1995}).
	
	\bibitem[{\citenamefont{Westra et~al.}(2010)\citenamefont{Westra, Poot, van~der
			Zant, and Venstra}}]{herreclamped}
	\bibinfo{author}{\bibfnamefont{H.~J.~R.} \bibnamefont{Westra}},
	\bibinfo{author}{\bibfnamefont{M.}~\bibnamefont{Poot}},
	\bibinfo{author}{\bibfnamefont{H.~S.~J.} \bibnamefont{van~der Zant}},
	\bibnamefont{and} \bibinfo{author}{\bibfnamefont{W.~J.}
		\bibnamefont{Venstra}}, \bibinfo{journal}{Phys. Rev. Lett.}
	\textbf{\bibinfo{volume}{105}}, \bibinfo{pages}{117205}
	(\bibinfo{year}{2010}).
	
	\bibitem[{\citenamefont{Antonio et~al.}(2012)\citenamefont{Antonio, Zanette,
			and L{\'o}pez}}]{antonio2012frequency}
	\bibinfo{author}{\bibfnamefont{D.}~\bibnamefont{Antonio}},
	\bibinfo{author}{\bibfnamefont{D.~H.} \bibnamefont{Zanette}},
	\bibnamefont{and}
	\bibinfo{author}{\bibfnamefont{D.}~\bibnamefont{L{\'o}pez}},
	\bibinfo{journal}{Nature communications} \textbf{\bibinfo{volume}{3}},
	\bibinfo{pages}{1} (\bibinfo{year}{2012}).
	
	\bibitem[{\citenamefont{Eichler et~al.}(2012)\citenamefont{Eichler, del
			{\'A}lamo~Ruiz, Plaza, and Bachtold}}]{eichler2012strong}
	\bibinfo{author}{\bibfnamefont{A.}~\bibnamefont{Eichler}},
	\bibinfo{author}{\bibfnamefont{M.}~\bibnamefont{del {\'A}lamo~Ruiz}},
	\bibinfo{author}{\bibfnamefont{J.}~\bibnamefont{Plaza}}, \bibnamefont{and}
	\bibinfo{author}{\bibfnamefont{A.}~\bibnamefont{Bachtold}},
	\bibinfo{journal}{Physical review letters} \textbf{\bibinfo{volume}{109}},
	\bibinfo{pages}{025503} (\bibinfo{year}{2012}).
	
	\bibitem[{\citenamefont{Chen et~al.}(2017)\citenamefont{Chen, Zanette,
			Czaplewski, Shaw, and L{\'o}pez}}]{chen2017direct}
	\bibinfo{author}{\bibfnamefont{C.}~\bibnamefont{Chen}},
	\bibinfo{author}{\bibfnamefont{D.~H.} \bibnamefont{Zanette}},
	\bibinfo{author}{\bibfnamefont{D.~A.} \bibnamefont{Czaplewski}},
	\bibinfo{author}{\bibfnamefont{S.}~\bibnamefont{Shaw}}, \bibnamefont{and}
	\bibinfo{author}{\bibfnamefont{D.}~\bibnamefont{L{\'o}pez}},
	\bibinfo{journal}{Nature communications} \textbf{\bibinfo{volume}{8}},
	\bibinfo{pages}{15523} (\bibinfo{year}{2017}).
	
	\bibitem[{\citenamefont{Shoshani et~al.}(2017)\citenamefont{Shoshani, Shaw, and
			Dykman}}]{shoshani2017anomalous}
	\bibinfo{author}{\bibfnamefont{O.}~\bibnamefont{Shoshani}},
	\bibinfo{author}{\bibfnamefont{S.~W.} \bibnamefont{Shaw}}, \bibnamefont{and}
	\bibinfo{author}{\bibfnamefont{M.~I.} \bibnamefont{Dykman}},
	\bibinfo{journal}{Scientific reports} \textbf{\bibinfo{volume}{7}},
	\bibinfo{pages}{18091} (\bibinfo{year}{2017}).
	
	\bibitem[{\citenamefont{Czaplewski et~al.}(2018)\citenamefont{Czaplewski, Chen,
			Lopez, Shoshani, Eriksson, Strachan, and Shaw}}]{czaplewski2018bifurcation}
	\bibinfo{author}{\bibfnamefont{D.~A.} \bibnamefont{Czaplewski}},
	\bibinfo{author}{\bibfnamefont{C.}~\bibnamefont{Chen}},
	\bibinfo{author}{\bibfnamefont{D.}~\bibnamefont{Lopez}},
	\bibinfo{author}{\bibfnamefont{O.}~\bibnamefont{Shoshani}},
	\bibinfo{author}{\bibfnamefont{A.~M.} \bibnamefont{Eriksson}},
	\bibinfo{author}{\bibfnamefont{S.}~\bibnamefont{Strachan}}, \bibnamefont{and}
	\bibinfo{author}{\bibfnamefont{S.~W.} \bibnamefont{Shaw}},
	\bibinfo{journal}{Physical review letters} \textbf{\bibinfo{volume}{121}},
	\bibinfo{pages}{244302} (\bibinfo{year}{2018}).
	
	\bibitem[{\citenamefont{Czaplewski et~al.}(2019)\citenamefont{Czaplewski,
			Strachan, Shoshani, Shaw, and L{\'o}pez}}]{czaplewski2019bifurcation}
	\bibinfo{author}{\bibfnamefont{D.~A.} \bibnamefont{Czaplewski}},
	\bibinfo{author}{\bibfnamefont{S.}~\bibnamefont{Strachan}},
	\bibinfo{author}{\bibfnamefont{O.}~\bibnamefont{Shoshani}},
	\bibinfo{author}{\bibfnamefont{S.~W.} \bibnamefont{Shaw}}, \bibnamefont{and}
	\bibinfo{author}{\bibfnamefont{D.}~\bibnamefont{L{\'o}pez}},
	\bibinfo{journal}{Applied Physics Letters} \textbf{\bibinfo{volume}{114}},
	\bibinfo{pages}{254104} (\bibinfo{year}{2019}).
	
	\bibitem[{\citenamefont{Yang et~al.}(2019)\citenamefont{Yang, Rochau, Huber,
			Brieussel, Rastelli, Weig, and Scheer}}]{yang2019spatial}
	\bibinfo{author}{\bibfnamefont{F.}~\bibnamefont{Yang}},
	\bibinfo{author}{\bibfnamefont{F.}~\bibnamefont{Rochau}},
	\bibinfo{author}{\bibfnamefont{J.~S.} \bibnamefont{Huber}},
	\bibinfo{author}{\bibfnamefont{A.}~\bibnamefont{Brieussel}},
	\bibinfo{author}{\bibfnamefont{G.}~\bibnamefont{Rastelli}},
	\bibinfo{author}{\bibfnamefont{E.~M.} \bibnamefont{Weig}}, \bibnamefont{and}
	\bibinfo{author}{\bibfnamefont{E.}~\bibnamefont{Scheer}},
	\bibinfo{journal}{Physical review letters} \textbf{\bibinfo{volume}{122}},
	\bibinfo{pages}{154301} (\bibinfo{year}{2019}).
	
	\bibitem[{\citenamefont{Houri et~al.}(2020)\citenamefont{Houri, Hatanaka,
			Asano, and Yamaguchi}}]{houri2019multimode}
	\bibinfo{author}{\bibfnamefont{S.}~\bibnamefont{Houri}},
	\bibinfo{author}{\bibfnamefont{D.}~\bibnamefont{Hatanaka}},
	\bibinfo{author}{\bibfnamefont{M.}~\bibnamefont{Asano}}, \bibnamefont{and}
	\bibinfo{author}{\bibfnamefont{H.}~\bibnamefont{Yamaguchi}},
	\bibinfo{journal}{Phys. Rev. Applied} \textbf{\bibinfo{volume}{13}},
	\bibinfo{pages}{014049} (\bibinfo{year}{2020}).
	
	\bibitem[{\citenamefont{Van~der Avoort et~al.}(2010)\citenamefont{Van~der
			Avoort, Van~der Hout, Bontemps, Steeneken, Le~Phan, Fey, Hulshof, and
			Van~Beek}}]{van2010amplitude}
	\bibinfo{author}{\bibfnamefont{C.}~\bibnamefont{Van~der Avoort}},
	\bibinfo{author}{\bibfnamefont{R.}~\bibnamefont{Van~der Hout}},
	\bibinfo{author}{\bibfnamefont{J.}~\bibnamefont{Bontemps}},
	\bibinfo{author}{\bibfnamefont{P.}~\bibnamefont{Steeneken}},
	\bibinfo{author}{\bibfnamefont{K.}~\bibnamefont{Le~Phan}},
	\bibinfo{author}{\bibfnamefont{R.}~\bibnamefont{Fey}},
	\bibinfo{author}{\bibfnamefont{J.}~\bibnamefont{Hulshof}}, \bibnamefont{and}
	\bibinfo{author}{\bibfnamefont{J.}~\bibnamefont{Van~Beek}},
	\bibinfo{journal}{Journal of Micromechanics and Microengineering}
	\textbf{\bibinfo{volume}{20}}, \bibinfo{pages}{105012}
	(\bibinfo{year}{2010}).
	
	\bibitem[{\citenamefont{Westra et~al.}(2011)\citenamefont{Westra, Karabacak,
			Brongersma, Crego-Calama, van~der Zant, and Venstra}}]{herredirectpar}
	\bibinfo{author}{\bibfnamefont{H.~J.~R.} \bibnamefont{Westra}},
	\bibinfo{author}{\bibfnamefont{D.~M.} \bibnamefont{Karabacak}},
	\bibinfo{author}{\bibfnamefont{S.~H.} \bibnamefont{Brongersma}},
	\bibinfo{author}{\bibfnamefont{M.}~\bibnamefont{Crego-Calama}},
	\bibinfo{author}{\bibfnamefont{H.~S.~J.} \bibnamefont{van~der Zant}},
	\bibnamefont{and} \bibinfo{author}{\bibfnamefont{W.~J.}
		\bibnamefont{Venstra}}, \bibinfo{journal}{Phys. Rev. B}
	\textbf{\bibinfo{volume}{84}}, \bibinfo{pages}{134305}
	(\bibinfo{year}{2011}).
	
	\bibitem[{\citenamefont{Dykman and Krivoglaz}(1975)}]{dykman1975spectral}
	\bibinfo{author}{\bibfnamefont{M.}~\bibnamefont{Dykman}} \bibnamefont{and}
	\bibinfo{author}{\bibfnamefont{M.}~\bibnamefont{Krivoglaz}},
	\bibinfo{journal}{physica status solidi (b)} \textbf{\bibinfo{volume}{68}},
	\bibinfo{pages}{111} (\bibinfo{year}{1975}).
	
	\bibitem[{\citenamefont{Dolleman et~al.}(2017)\citenamefont{Dolleman,
			Davidovikj, van~der Zant, and Steeneken}}]{robincalib}
	\bibinfo{author}{\bibfnamefont{R.~J.} \bibnamefont{Dolleman}},
	\bibinfo{author}{\bibfnamefont{D.}~\bibnamefont{Davidovikj}},
	\bibinfo{author}{\bibfnamefont{H.~S.~J.} \bibnamefont{van~der Zant}},
	\bibnamefont{and} \bibinfo{author}{\bibfnamefont{P.~G.}
		\bibnamefont{Steeneken}}, \bibinfo{journal}{Applied Physics Letters}
	\textbf{\bibinfo{volume}{111}}, \bibinfo{pages}{253104}
	(\bibinfo{year}{2017}).
	
	\bibitem[{SM()}]{SM}
	\bibinfo{note}{See Supplemental Material at http://link.aps.org/, which
		includes a description of the measuremnts calibration, derivation of the
		equations of motion, model analysis, and model calibration}.
	
	\bibitem[{\citenamefont{Lifshitz and Cross}(2008)}]{lifshitz2008nonlinear}
	\bibinfo{author}{\bibfnamefont{R.}~\bibnamefont{Lifshitz}} \bibnamefont{and}
	\bibinfo{author}{\bibfnamefont{M.}~\bibnamefont{Cross}},
	\bibinfo{journal}{Review of nonlinear dynamics and complexity}
	\textbf{\bibinfo{volume}{1}}, \bibinfo{pages}{1} (\bibinfo{year}{2008}).
	
	\bibitem[{\citenamefont{Dolleman et~al.}(2018)\citenamefont{Dolleman, Houri,
			Chandrashekar, Alijani, van~der Zant, and Steeneken}}]{robinparam}
	\bibinfo{author}{\bibfnamefont{R.~J.} \bibnamefont{Dolleman}},
	\bibinfo{author}{\bibfnamefont{S.}~\bibnamefont{Houri}},
	\bibinfo{author}{\bibfnamefont{A.}~\bibnamefont{Chandrashekar}},
	\bibinfo{author}{\bibfnamefont{F.}~\bibnamefont{Alijani}},
	\bibinfo{author}{\bibfnamefont{H.~S.~J.} \bibnamefont{van~der Zant}},
	\bibnamefont{and} \bibinfo{author}{\bibfnamefont{P.~G.}
		\bibnamefont{Steeneken}}, \bibinfo{journal}{Scientific Reports}
	\textbf{\bibinfo{volume}{8}}, \bibinfo{pages}{9366} (\bibinfo{year}{2018}).
	
\end{thebibliography}
\end{document}